\documentclass[11pt]{article}
\usepackage[letterpaper,hmargin=1in,vmargin=1.25in]{geometry}

\usepackage{fullpage}
\usepackage{amssymb}
\usepackage{amsthm}
\usepackage{graphicx}
\usepackage{wrapfig}
\usepackage{caption,subcaption}
\usepackage{amsmath,amssymb,stmaryrd}
\usepackage{url}  
\usepackage{algorithm}
\usepackage[noend]{algpseudocode}
\usepackage{sidecap}

\newcommand{\Link}[2]{\mathbf{Link}\paren{#1,#2}}
\newcommand{\Recall}[1]{\mathbf{Recall}\paren{#1}}
\newcommand{\ReuseWeakPassword}{{\bf Reuse Weak}}
\newcommand{\ReuseStrongPassword}{{\bf Reuse Strong}}
\newcommand{\Lifehacker}{{\bf Lifehacker}}
\newcommand{\StrongRandomPassword}{{\bf Strong Random and Independent}} 

\newcommand{\Generator}{\mathcal{G}}

\newcommand{\sharedCues}{Shared Cues}
\newcommand{\ExtraRehearsals}[2]{X_{#1,#2}}
\newcommand{\TotalExtraRehearsals}[1]{X_{#1}}

\newcommand{\PasswordSpace}{\mathcal{P}}
\newcommand{\CueSpace}{\mathcal{C}}
\newcommand{\AssocSpace}{\mathcal{AS}}
\newcommand{\UserKnowledge}{\mathcal{K}}

\newcommand{\Guesses}[1]{q_{#1}}
\newcommand{\Strikes}{s}
\newcommand{\NumPhish}{r}
\newcommand{\NumHashes}{h}
\newcommand{\NumBaseCues}{n}
\newcommand{\AssociationStrength}[1]{\sigma_{#1}}
\newcommand{\Adversary}{\mathcal{A}}
\newcommand{\User}{\mathcal{U}}
\newcommand{\Action}{\mathcal{ACT}}
\newcommand{\Object}{\mathcal{OBJ}}
\newcommand{\Entropy}{H}
\newcommand{\MinEntropy}{H_{\min}}
\newcommand{\MinConditionalEntropy}{HC_{\min }}
\newcommand{\MinConditionalKDeltaEntropy}[2]{HC_{\min } \paren{#1,#2}}
\newcommand{\paren}[1]{\left( #1 \right)}
\newcommand{\cut}[1]{}
\newtheorem{definition}{Definition}
\newtheorem{claim}{Claim}
\newtheorem{theorem}{Theorem}
\newtheorem{lemma}{Lemma}

\def \QED {\hfill{$\Box$}}

\newenvironment{proofof}[1]{\noindent {\em Proof of #1.  }}{\QED}

\newenvironment{remindertheorem}[1]{\medskip \noindent {\bf Reminder of Theorem #1.  }\em}{}

\newenvironment{reminderlemma}[1]{\medskip \noindent {\bf Reminder of Lemma #1.  }\em}{}

\begin{document}

%\mainmatter  % start of an individual contribution

\title{Naturally Rehearsing Passwords}

%\titlerunning{Self Rehearsing Passwords}

\author{Jeremiah Blocki \thanks{This work was partially supported by the NSF Science and Technology TRUST and the AFOSR MURI on Science of Cybersecurity. The first author was also partially supported by an NSF Graduate Fellowship.} \\ 
Carnegie Mellon University\\ 
5000 Forbes Avenue, Pittsburgh, PA. \\
 \and Manuel Blum \\ 
Carnegie Mellon University \\ 
5000 Forbes Avenue, Pittsburgh, PA. \\
 \and Anupam Datta   \\ 
Carnegie Mellon University\\
5000 Forbes Avenue, Pittsburgh, PA.\\
}
%\authorrunning{Anonymous Submission}

%\institute{}


%\toctitle{Self-Rehearsing Passwords}
%\tocauthor{Anonymous}
\maketitle

\begin{abstract}
We introduce quantitative usability and security models to guide the design of \emph{password 
management schemes} ---  systematic strategies to help users create and remember multiple 
passwords. In the same way that security proofs in cryptography are based on  
complexity-theoretic assumptions (e.g., hardness of factoring and discrete logarithm), we quantify 
usability by introducing \emph{usability assumptions}. In particular, password management relies 
on assumptions about human memory, e.g., that a user who follows a particular rehearsal 
schedule will successfully maintain the corresponding memory. These assumptions are informed by research in cognitive science and can be tested empirically.  Given rehearsal requirements and a user's 
visitation schedule for each account, we use the total number of extra rehearsals that 
the user would have to do to remember all of his passwords as a measure of the usability of 
the password scheme. Our usability model leads us to a key observation: password reuse benefits users not only by reducing the number of passwords that the user has to memorize, but more importantly by increasing the natural rehearsal rate for each password.  We also present a security model which accounts for the complexity of password 
management with multiple accounts and associated threats, 
including online, offline, and plaintext password leak attacks. Observing that current 
password management schemes are either insecure or unusable, we present 
\sharedCues~--- a new scheme in which the underlying secret is strategically 
shared across accounts to ensure that most rehearsal requirements are satisfied naturally while  
simultaneously providing strong security. The construction uses the Chinese Remainder Theorem to achieve these competing goals. 
\end{abstract}


\section{Introduction}\label{sec:introduction} 
A typical computer user today manages passwords for many 
different online accounts. 
Users struggle with this task---often forgetting their passwords or 
adopting insecure practices, such as using the same 
password for multiple accounts and selecting weak 
passwords~\cite{florencio2007large,center2010consumer,kruger2008empirical,bonneau2012science}.
While there are many articles, books, papers and even comics 
about selecting strong individual passwords
\cite{burnett2005perfect,XKCDhorsebatterystaplecorrect,Gaw:2006:PMS:1143120.1143127,yan2004password,timePimpMyPassword,guideline:DOD1985,guideline:NIST2009,guideline:lifehacker}, 
there is very little work on \emph{password management schemes}---systematic 
strategies to help users create and remember multiple passwords---that are both usable and
secure. In this paper, we present a rigorous treatment of password management schemes. 
Our contributions include a formalization of important aspects of a usable scheme,  
a quantitative security model, and a construction that provably achieves the competing security and 
usability properties. 

\paragraph{Usability Challenge.} We consider a setting where a user has two 
types of memory: {\em persistent memory} (e.g., a sticky note or a text file on
his computer) and {\em associative memory} (e.g., his own human memory). We assume that
persistent memory is reliable and convenient but not private (i.e., accessible to an 
adversary). In contrast, a user's associative memory is private but lossy---if the user does not
rehearse a memory it may be forgotten. 
While our understanding of human memory is incomplete, it has
been an active area of research \cite{memory:textbook:baddeley1997} and there
are many mathematical models of human memory
\cite{memory:AssociativeSystemTheoretical:kohonen1977associative,memory:MarrAssessment:willshaw1990,memory:act-r:anderson1997act,memory:marr1971,valiant2005memorization}.
These models differ in many details, but they all model an associative memory
with cue-association pairs: to remember $\hat{a}$ (e.g., a password) the brain associates the memory
with a context $\hat{c}$ (e.g., a public hint or cue); such associations are strengthened by 
rehearsal \cut{\footnote{Physically, the cue-association pair $(\hat{c},\hat{a})$ might encode
the excitement levels of the neurons in the user's brain\cite{memory:marr1971}.}}.
A central challenge in designing usable password schemes is thus to create associations
that are strong and to maintain them over time through rehearsal. Ideally, we would 
like the rehearsals to be \emph{natural}, i.e., they should be a side-effect of users'
normal online activity. Indeed insecure password management practices adopted by users, 
such as reusing passwords, improve usability by increasing the number of times a password is 
naturally rehearsed as users visit their online accounts. 
%Florencio and Herley estimate that 1.5\% of Yahoo users forget their password every month
%\cite{florencio2007large}. It is especially challenging to remember a
%complicated password that is rehearsed infrequently. Many users have nearly
%one-hundred different online accounts --- most of which are visited
%infrequently. Intimidated by the prospect of creating and rehearsing so many
%different passwords, users often adopt insecure password practices: writing
%down passwords, reusing passwords and picking weak passwords which are
%vulnerable to password cracking attacks. From a usability standpoint, the
%practice of password reuse minimizes the total number of passwords that the
%user has to remember and helps the user to rehearse his passwords naturally. 

\paragraph{Security Challenge.} Secure password management is not merely a
theoretical problem---there are numerous real-world examples of password
breaches
\cite{breach:CERT-Warning,center2010consumer,breach:militaryHACK,breach:natoHACK,noPlaintextPassword,breach:Zappos,breach:Atlassian,breach:apple,breach:sony,breach:linkedin,breach:IEEE}.
Adversaries may crack a weak password in an \emph{online attack} where they simply 
visit the online account and try as many guesses as the site permits. In many cases (e.g.,
Zappos, LinkedIn, Sony, Gawker
\cite{breach:Zappos,breach:Atlassian,breach:sony,breach:natoHACK,breach:militaryHACK,breach:linkedin})
an adversary is able to mount an \emph{offline attack} to crack weak
passwords after the cryptographic hash of a password is leaked or stolen. To
protect against an offline attack, users are often advised to pick long
passwords that include numbers, special characters and capital letters
\cite{guideline:NIST2009}. In other cases even the strongest passwords are
compromised  via a \emph{plaintext password leak attack} (e.g.,
\cite{breach:rockyou,breach:apple,noPlaintextPassword,breach:IEEE}), for example, because
the user fell prey to a phishing attack or signed into his
account on an infected computer or because of server misconfigurations. Consequently, users are typically advised
against reusing the same password. A secure password management scheme must
protect against all these types of breaches. 

\paragraph{Contributions.} We precisely define the password management problem
in Section~\ref{sec:Preliminaries}. A password management scheme consists of 
a \emph{generator}---a function that outputs a set of public cue-password 
pairs---and a \emph{rehearsal schedule}. The generator is implemented using a 
computer program whereas the human user is expected to follow the rehearsal 
schedule for each cue. This division of work is critical---the computer program performs
tasks that are difficult for human users (e.g., generating random bits) whereas 
the human user's associative memory is used to store passwords since the computer's
persistent memory is accessible to the adversary. \\

\noindent{\it Quantifying Usability.} 
In the same way that security proofs in cryptography are based on  
complexity-theoretic assumptions (e.g., hardness of factoring and discrete logarithm), we quantify 
usability by introducing \emph{usability assumptions}. In particular, password management relies 
on assumptions about human memory, e.g., that a user who follows a particular rehearsal 
schedule will successfully maintain the corresponding memory. These assumptions are informed by research in cognitive science and can be tested empirically. Given rehearsal requirements and a user's 
visitation schedule for each account, we use the total number of extra rehearsals that 
the user would have to do to remember all of his passwords as a measure of the usability of 
the password scheme (Section \ref{sec:usability}). Specifically, in our usability analysis, we use the 
\emph{Expanding Rehearsal Assumption (ER)} that allows
for memories to be rehearsed with exponentially decreasing frequency, i.e., 
rehearse at least once in the time-intervals (days) $\left[1,2\right)$, $\left[2,4\right)$, 
$\left[4,8\right)$ and so on. Few long-term memory experiments have been conducted, but \emph{ER} is
consistent with known studies \cite{memory:forgetting:squire1989,memory:ExpandingRehearsal}. Our memory
assumptions are parameterized by a constant $\AssociationStrength{}$ which
represents the strength of the mnemonic devices used to memorize and rehearse a
cue-association pair. Strong mnemonic 
techniques~\cite{memory:memoryPalaceMatteoRicci:spence1985,foer2011moonwalking}
exploit the associative nature of human memory discussed earlier and its 
remarkable visual/spatial capacity~\cite{Memory:10000Pictures:standingt1973}. \\

\noindent {\it Quantifying Security.}  We present a game based security model
for a password management scheme (Section \ref{sec:Security}) in the style
of exact security definitions~\cite{bellare1996exact}. The game is played 
between a user ($\User$) and a
resource-bounded adversary ($\Adversary$) whose goal is to guess one of the
user's passwords. Our game models three commonly occurring breaches (online
attack, offline attack, plaintext password leak attack).  \\

\noindent{\it Our Construction.} We present a new password management scheme,
which we call  \sharedCues, and prove that it provides strong security and usability 
properties (see Section \ref{sec:PicturesAsCues}). Our scheme
incorporates powerful mnemonic techniques through the use of public cues (e.g.,
photos) to create strong associations. The user first associates a randomly
generated person-action-object story (e.g., Bill Gates swallowing a bike) with
each public cue. We use the Chinese Remainder Theorem to share cues across sites in a way that balances
several competing security and usability goals:
 1) Each cue-association pair is used by many different web sites (so
that most rehearsal requirements are satisfied naturally), 2) the total number 
of cue-association pairs that the user has to memorize is low, 3) each web
site  uses several cue-association pairs (so that passwords are secure)
and 4) no two web sites share too many cues (so that passwords remain
secure even after the adversary obtains some of the user's other passwords). We show that our construction achieves an asymptotically optimal balance between these security and usability goals (Lemma \ref{lemma:Intersection}, Theorem \ref{thm:SecurityUpperBound}).

\paragraph{Related Work.} %\label{subsec:related}
A distinctive goal of our work is to quantify usability of password management schemes 
by drawing on ideas from cognitive science and leverage this understanding to design 
schemes with acceptable usability. We view the results of this paper--employing usability
assumptions about rehearsal requirements---as an initial step towards this goal. 
While the mathematical constructions start from the usability assumptions, the 
assumptions themselves are empirically testable, e.g., via longitudinal user studies.
In contrast, a line of prior work on usability has focused on empirical 
studies of user behavior including their password management habits~\cite{florencio2007large,center2010consumer,kruger2008empirical},
the effects of password composition rules (e.g., requiring numbers and special symbols) on individual passwords
\cite{usability:compositionPolicies,blocki2013optimizing}, the memorability of individual system assigned passwords
\cite{usabilitystudy:xkcd}, graphical passwords~\cite{brostoff2000passfaces,biddle2012graphical}, 
and passwords based on implicit learning~\cite{rubberHose}. These user studies
have been limited in duration and scope (e.g., study retention of a single
password over a short period of time). Other work~\cite{bonneau2012quest} 
articulates informal, but more comprehensive, usability criteria for password schemes.

Our use of cued recall is driven by evidence that it is much easier than pure recall
\cite{memory:textbook:baddeley1997}. We also exploit the large human capacity for visual
memory~\cite{Memory:10000Pictures:standingt1973} by using pictures as cues. 
Prior work on graphical passwords
\cite{brostoff2000passfaces,biddle2012graphical} also takes advantage of these features.
However, our work is distinct from the literature on graphical passwords 
because we address the challenge of managing multiple passwords. More generally, 
usable and secure password management is an excellent problem  to explore deeper connections between cryptography and cognitive science.
%---our work as well as the empirical work on graphical 
%and implicit-learning based passwords can be viewed as initial results in this 
%exploration~\cite{rubberHose}.

Security metrics for passwords like (partial) guessing entropy (e.g., how many
guesses does the adversary need to crack $\alpha$-fraction of the passwords in
a dataset
\cite{massey1994guessing,pliam2000incomparability,bonneau2012science}? how many
passwords can the adversary break with $\beta$ guesses per account
\cite{boztas1999entropies}?) were designed to analyze the security of a dataset
of passwords from many users, not the security of a particular user's password management scheme. While these
metrics can provide useful feedback about individual passwords (e.g., they rule out some insecure passwords) they do not deal with the complexities of securing multiple accounts against an adversary who may have gained background knowledge about the user from previous attacks --- we refer an interested reader to the full version \cite{fullVersion} of this paper for more discussion.
 
Our notion of $(n,\ell,\gamma)$-sharing set families (definition \ref{def:GoodSharing}) is equivalent to Nisan and Widgerson's definition of a $(k,m)$-design \cite{setSharing}. However, Nisan and Widgerson were focused on a different application (constructing pseudorandom bit generators) and the range of parameters that they consider are not suitable for our password setting in which $\ell$ and $\gamma$ are constants. See the full version\cite{fullVersion} of this paper for more discussion.   

\cut{ }

\section{Definitions} \label{sec:Preliminaries} 
We use $\PasswordSpace$ to denote the space of possible passwords. A password management scheme needs to generate $m$ passwords $p_1,...,p_m \in \PasswordSpace$ --- one for each account $A_i$. %, and we use $k \in \UserKnowledge$ to denote any knowledge that the user has (e.g., birth dates, names of friends, familiar songs)

\paragraph{Associative Memory and Cue-Association Pairs.} Human memory is associative. Competitors in memory competitions routinely use mnemonic techniques (e.g., the method of loci \cite{memory:memoryPalaceMatteoRicci:spence1985}) which exploit associative memory\cite{foer2011moonwalking}. For example, to remember the word `apple' a competitor might imagine a giant apple on the floor in his bedroom. The bedroom now provides a context which can later be used as a cue to help the competitor remember the word apple. We use $\hat{c} \in \CueSpace$ to denote the cue, and we use $\hat{a} \in \AssocSpace$ to denote the corresponding association in a cue-association pair $\paren{\hat{c},\hat{a}}$. Physically, $\hat{c}$ (resp. $\hat{a}$) might encode the excitement levels of the neurons in the user's brain when he thinks about his bedroom (resp. apples) \cite{memory:marr1971}.  
    
We allow the password management scheme to store $m$ sets of public cues $c_1,...,c_m \subset \CueSpace$ in persistent memory to help the user remember each password. Because these cues are stored in persistent memory they are always available to the adversary as well as the user. Notice that a password may be derived from multiple cue-association pairs. We use $\hat{c} \in \mathcal{C}$ to denote a cue, $c \subset \CueSpace$ to denote a set of cues, and $C = \bigcup_{i=1}^m c_i$ to denote the set of all cues --- $\NumBaseCues = \left| C\right|$ denotes the total number of cue-association pairs that the user has to remember.

\paragraph{Visitation Schedules and Rehearsal Requirements.} Each cue $\hat{c} \in
C$ may have a rehearsal schedule to ensure that the cue-association
pair $(\hat{c},\hat{a})$ is maintained.  \begin{definition}
\label{def:RehearsalRequirement} A rehearsal schedule for a
cue-association pair $(\hat{c},\hat{a})$ is a sequence of times $t_0^{\hat{c}} < t_1^{\hat{c}} <...$. For each $i \geq 0$ we have a {\em rehearsal requirement}, the cue-association pair must be rehearsed at least once during the time window $\left[t_i^{\hat{c}},t_{i+1}^{\hat{c}}\right) = \{x \in \mathbb{R} ~ \vline~
t_i^{\hat{c}} \leq x < t_{i+1}^{\hat{c}} \}$.  \end{definition} A rehearsal schedule is {\em
sufficient} if a user can maintain the association $\left(\hat{c},\hat{a}\right)$ by following the
rehearsal schedule. We discuss sufficient rehearsal assumptions in section
\ref{sec:usability}. The length of each
interval $\left[t_{i}^{\hat{c}},t_{i+1}^{\hat{c}}\right)$ may depend on the strength of the
mnemonic technique used to memorize and rehearse a cue-association pair $\paren{\hat{c},\hat{a}}$
as well as $i$ --- the number of prior rehearsals. For notational convenience, we use a function $R: C\times \mathbb{N} \rightarrow \mathbb{R}$ to specify the rehearsal requirements   (e.g., $R\paren{\hat{c},j} = t_j^{\hat{c}}$), and we use $\mathcal{R}$ to denote a set of rehearsal functions.

A visitation schedule for an account $A_i$ is a sequence of real numbers
$\tau_0^i < \tau_1^i < \ldots$, which represent the times when the account
$A_i$ is visited by the user. We do not assume that the exact visitation
schedules are known a priori. Instead we model visitation schedules using a
random process with a known parameter $\lambda_i$ based on
$E\left[\tau_{j+1}^i-\tau_j^i\right]$ --- the average time between consecutive
visits to account $A_i$. A rehearsal requirement $\left[t_i^{\hat{c}},t_{i+1}^{\hat{c}}\right)$
can be satisfied naturally if the user visits a site $A_j$ that
uses the cue $\hat{c}$ $\paren{\hat{c} \in c_j}$  during the given time window. Formally,

\begin{definition} We say that a rehearsal requirement
$\left[t_i^{\hat{c}},t_{i+1}^{\hat{c}}\right)$ is {\em naturally satisfied} by a visitation
schedule $\tau_0^i < \tau_1^i < \ldots$ if $\exists j \in [m],k \in \mathbb{N}$ s.t $\hat{c} \in c_j$ and $\tau_k^j
\in \left[t_i^{\hat{c}},t_{i+1}^{\hat{c}}\right)$.  We use 
\[\ExtraRehearsals{t}{\hat{c}} = \left|\left\{i ~\vline ~ t_{i+1}^{\hat{c}}
\leq t \wedge \forall j,k. \left(\hat{c} \notin c_j \vee \tau_k^j \notin
\left[t_i^{\hat{c}},t_{i+1}^{\hat{c}} \right) \right) \right\} \right| \ , \]
to denote the number of rehearsal requirements that are not naturally satisfied by the visitation schedule during the time interval $[0,t]$.
\end{definition}

We use rehearsal requirements and visitation schedules to quantify the
usability of a password management scheme by measuring the total number of
extra rehearsals. If a cue-association pair $\left(\hat{c},\hat{a}\right)$ is not rehearsed naturally during the interval $\left[t^{\hat{c}}_i,t^{\hat{c}}_{i+1}\right)$ then the user needs to perform an extra rehearsal to maintain the association. Intuitively, $\ExtraRehearsals{t}{\hat{c}}$ denotes the total number of extra rehearsals of the cue-association pair $\left(\hat{c},\hat{a}\right)$ during the time interval $[0,t]$. We use $\TotalExtraRehearsals{t} = \sum_{\hat{c} \in C}
\ExtraRehearsals{t}{\hat{c}}$ to denote the total number of extra rehearsals during the time interval $[0,t]$ to maintain all of the cue-assocation pairs. \\

{\noindent\bf Usability Goal: } Minimize the expected value of
$E\left[\TotalExtraRehearsals{t}\right]$.

\paragraph{Password Management Scheme.} A password management scheme includes a
generator $\Generator_m$ and a rehearsal schedule $R \in \mathcal{R}$. The generator $\Generator_m\paren{k,b,\vec{\lambda},R}$ utilizes a user's
knowledge $k \in \UserKnowledge$, random bits $b \in \{0,1\}^*$ to generate passwords
$p_1,...,p_m$ and public cues $c_1,...,c_m \subseteq \mathcal{C}$. $\Generator_m$ may use the rehearsal schedule $R$ and the visitation schedules $\vec{\lambda} = \langle\lambda_1,...,\lambda_m \rangle$ of each site to help minimize $E\left[\TotalExtraRehearsals{t}\right]$. Because the
cues $c_1,...c_m$ are public they may be stored in persistent memory along with
the code for the generator $\Generator_m$. In contrast, the passwords
$p_1,...p_m$ must be memorized and rehearsed by the user (following $R$) so that the cue association pairs $(c_i,p_i)$ are maintained in his associative memory.

\begin{definition} A password management scheme is a tuple $\langle
\Generator_m, R \rangle$, where $\Generator_m$ is a function $\Generator_m:
\UserKnowledge \times \{0,1\}^* \times \mathbb{R}^m \times \mathcal{R} \rightarrow
\paren{\PasswordSpace \times 2^\CueSpace}^m$ and a $R \in \mathcal{R}$ is a rehearsal schedule
which the user must follow for each cue.  \end{definition}

Our security analysis is not based on the secrecy of
$\Generator_m$, $k$ or the public cues $C = \bigcup_{i=1}^m c_i$. The adversary will be able to find the cues $c_1,...,c_m$ because they are stored in persistent memory. In fact, we also assume that the adversary has background knowledge about the user (e.g., he may know $k$), and  that the adversary knows the password management scheme
$\Generator_m$.  The only secret is the random string $b$ used by
$\Generator_m$ to produce $p_1,...,p_m$.  \\
\noindent{\bf Example Password Management Schemes.~}  Most password suggestions are too vague (e.g.,``pick an obscure phrase that is personally meaningful to you") to satisfy the precise requirements of a password management scheme --- formal security proofs of protocols involving human interaction can break down when humans behave in unexpected ways due to vague instructions \cite{radke2012towards}. We
consider the following formalization of password management schemes: (1)
\ReuseWeakPassword~--- the user selects a random dictionary word $w$ (e.g.,
from a dictionary of $20,000$ words) and uses  $p_i = w$ as the password for
every account $A_i$. (2) \ReuseStrongPassword~--- the user selects four random
dictionary words ($w_1,w_2,w_3,w_4$) and uses $p_i=w_1w_2w_3w_4$ as the
password for every account $A_i$. (3) \Lifehacker~(e.g.,
\cite{guideline:lifehacker}) --- The user selects three random words
($w_1,w_2,w_3$) from the dictionary as a base password $b = w_1w_2w_3$. The
user also selects a random derivation rule $d$ to derive a string from each
account name (e.g., use the first three letters of the account name, use the
first three vowels in the account name). The password for account $A_i$ is $p_i
= bd\left(A_i\right)$ where $d\left(A_i\right)$ denotes the derived string. (4)
\StrongRandomPassword~--- for each account $A_i$ the user selects four fresh
words independently at random from the dictionary and uses $p_i =
w_1^iw_2^iw_3^iw_4^i$. Schemes (1)-(3) are formalizations of popular password
management strategies. We argue that they are popular because they are easy to
use, while the strongly secure scheme \StrongRandomPassword~is unpopular
because the user must spend a lot of extra time rehearsing his passwords. See
the full version \cite{fullVersion} of this paper for more discussion of the security and usability of each scheme.

% don't guide the design of pms, user studies can help us evaluate candidate schemes
\section{Usability Model} \label{sec:usability}
People typically adopt their password management scheme based on usability considerations instead of security considerations \cite{florencio2007large}. Our usability model can be used to explain why users tend to adopt insecure password management schemes like \ReuseWeakPassword, \Lifehacker, or \ReuseStrongPassword. Our usability metric measures the extra effort that a user has to spend rehearsing his passwords. Our measurement depends on three important factors: rehearsal requirements for each cue, visitation rates for each site, and the total number of cues that the user needs to maintain. Our main technical result in this section is Theorem \ref{thm:ExtraRehearsals} --- a formula to compute the total number of extra rehearsals that a user has to do to maintain all of his passwords for $t$ days. To evaluate the formula we need to know the rehearsal requirements for each cue-association pair as well as the visitation frequency $\lambda_i$ for each account $A_i$. 

\paragraph{Rehearsal Requirements.} \label{subsec:rehearsalRequirements}
If the password management scheme does not mandate sufficient rehearsal then the user might forget his passwords. Few memory studies have attempted to study memory retention over long periods of time so we do not know exactly what these rehearsal constraints should look like. While security proofs in cryptography are based on assumptions from complexity theory (e.g., hardness of factoring and discrete logarithm), we need to make assumptions about humans. For example, the assumption behind CAPTCHAs is that humans are able to perform a simple task like reading garbled text \cite{captcha}. A rehearsal assumption specifies what types of rehearsal constraints are sufficient to maintain a memory. We consider two different assumptions about sufficient rehearsal schedules: Constant Rehearsal Assumption (CR) and Expanding Rehearsal Assumption (ER). Because some mnemonic devices are more effective than others (e.g., many people have amazing visual and spatial memories \cite{Memory:10000Pictures:standingt1973}) our assumptions are parameterized by a constant $\AssociationStrength{}$ which represents the strength of the mnemonic devices used to memorize and rehearse a cue association  pair. \\

{\bf Constant Rehearsal Assumption (CR):} The rehearsal schedule given by $R\paren{\hat{c},i} = i\AssociationStrength{}$ is sufficient to maintain the association $(\hat{c},\hat{a})$.  \\

CR is a pessimistic assumption --- it asserts that memories are not permanently strengthened by rehearsal. The user must continue rehearsing every $\AssociationStrength{}$ days --- even if the user has frequently rehearsed the password in the past. 

{\bf Expanding Rehearsal Assumption (ER):} The rehearsal schedule given by $R\paren{\hat{c},i} = 2^{i\AssociationStrength{}}$ is sufficient to maintain the association $(\hat{c},\hat{a})$.  \\ 

ER is more optimistic than CR --- it asserts that memories are strengthened by rehearsal so that memories need to be rehearsed less and less frequently as time passes. If a password has already been rehearsed $i$ times then the user does not have to rehearse again for $2^{i\AssociationStrength{}}$ days to satisfy the rehearsal requirement $\left[2^{i\AssociationStrength{}}, 2^{i\AssociationStrength{}+\AssociationStrength{}} \right)$. ER is  consistent with several long term memory experiments \cite{memory:forgetting:squire1989},\cite[Chapter 7]{memory:textbook:baddeley1997}, \cite{memory:ExpandingRehearsal} --- we refer the interested reader to full version\cite{fullVersion} of this paper for more discussion. We also consider the rehearsal schedule $R\paren{\hat{c},i} = i^2$ (derived from \cite{memory:alternateanderson1991reflections,memory:alternatevan}) in the full version ---   the usability results are almost indentical to those for ER.

\paragraph{Visitation Schedules.}% \label{subsec:VisitationSchedules}
Visitation schedules may vary greatly from person to person. For example, a 2006 survey about Facebook usage showed that $47\%$ of users logged in daily,  $22.4\%$ logged in about twice a week, $8.6\%$ logged in about once a week, and $12\%$ logged in about once a month\cite{facebookLoginFrequency}. We use a Poisson process with parameter $\lambda_i$ to model the visitation schedule for site $A_i$. We assume that the value of $1/\lambda_i$ --- the average inter-visitation time --- is known. For example, some websites (e.g., gmail) may be visited daily ($\lambda_i=1/1$ day) while other websites (e.g., IRS) may only be visited once a year on average (e.g., $\lambda_i = 1/365$ days). The Poisson process has been used to model the distribution of requests to a web server \cite{rasch1963poisson}. While the Poisson process certainly does not perfectly model a user's visitation schedule (e.g., visits to the IRS websites may be seasonal) we believe that the predictions we derive using this model will still be useful in guiding the development of usable password management schemes. While we focus on the Poisson arrival process, our analysis could be repeated for other random processes.

We consider four very different types of internet users: very active, typical, occasional and infrequent. Each user account $A_i$ may be visited daily (e.g., $\lambda_i = 1$), every three days  ($\lambda_i = 1/3$), every week (e.g. $\lambda_i = 1/7$), monthly $(\lambda_i = 1/31)$, or yearly $(\lambda_i=1/365)$ on average. See table \ref{tab:userSchedules} to see the full visitation schedules we define for each type of user. For example, our very active user has $10$ accounts he visits daily and $35$ accounts he visits annually. %Comment on the need for studies to inform real life rehearsal schedules. We can find data for individual sites like facebook, but we need corellated data.

\begin{table}[t]
\parbox{.40\linewidth}{
\centering
\begin{tabular}{| l | c | c | c | c | c |}
\hline
Schedule $\vline~ \lambda$& $\frac{1}{1 }$  & $\frac{1}{3 }$  & $\frac{1}{7}$  & $\frac{1}{31 }$  & $\frac{1}{365 }$ \\
\hline
Very Active & 10 & 10 & 10 & 10 & 35 \\
\hline
Typical & 5 & 10 & 10 & 10 & 40 \\
\hline
Occasional & 2 & 10 & 20 & 20 & 23 \\
\hline
Infrequent & 0 & 2 & 5 & 10 & 58 \\
\hline 
\end{tabular}
\caption{Visitation Schedules - number of accounts visited with frequency $\lambda$ (visits/days)}
\label{tab:userSchedules}
%\end{table}
}
\hfill 
%\begin{table}[h]
%\centering
\parbox{.55\linewidth}{
\begin{tabular}{| l | l | l || l |l  | }
\hline 
Assumption & \multicolumn{2}{c|}{CR $(\AssociationStrength{}=1)$ } & \multicolumn{2}{||c|}{ER $(\AssociationStrength{}=1)$} \\
\hline
Schedule/Scheme  & B+D & SRI &  B+D & SRI  \\
\hline
Very Active & $\approx 0$ & $23,396$ & $.023$ & $420$ \\
\hline
Typical & $.014$ & $24,545$  & $.084$ & $456.6$  \\
\hline 
Occasional & .$05$ & $24,652$ & $.12$ & $502.7$ \\
\hline
Infrequent & $56.7$ & $26,751$ & $1.2$ & $564$  \\
\hline
\end{tabular}
\caption{$E\left[\TotalExtraRehearsals{365}\right]$: Extra Rehearsals over the first year for both rehearsal assumptions. \newline B+D: \Lifehacker ~~~ \newline SRI: \StrongRandomPassword } 
\label{tab:UsabilityOld}
}
 \hfill
%\vspace{-.3in}
\end{table}

\paragraph{Extra Rehearsals.} Theorem \ref{thm:ExtraRehearsals} leads us to our key observation: cue-sharing benefits users both by (1) reducing the number of cue-association pairs that the user has to memorize and (2) by increasing the rate of natural rehearsals for each cue-association pair. For example, a active user with $75$ accounts  would need to perform $420$ extra-rehearsals over the first year to satisfy the rehearsal requirements given by ER if he adopts \StrongRandomPassword~ or just $0.023$ with \Lifehacker~--- see table \ref{tab:UsabilityOld}. The number of unique cue-association pairs $n$ decreased by a factor of $75$, but the total number of extra rehearsals $E[\TotalExtraRehearsals{365}]$ decreased by a factor of $8,260.8 \approx 75\times 243$ due to the increased natural rehearsal rate.

\newcommand{\thmExtraRehearsals}{Let $i_{\hat{c}}* = \left(\arg\max_x t^{\hat{c}}_{x} < t \right)-1$ then \begin{eqnarray*}
E\left[\TotalExtraRehearsals{t} \right] = \sum_{\hat{c} \in C} \sum_{i=0}^{i_{\hat{c}}*} \exp \left(-\left(\sum_{j:\hat{c} \in c_j} \lambda_j \right)\left(t^{\hat{c}}_{i+1}-t^{\hat{c}}_i \right) \right) \ 
\end{eqnarray*}  

}

\begin{theorem} \label{thm:ExtraRehearsals}
\thmExtraRehearsals
\end{theorem}

Theorem \ref{thm:ExtraRehearsals} follows easily from Lemma \ref{lemma:ExponentialDistribution} and linearity of expectations. Each cue-association pair $\paren{\hat{c},\hat{a}}$ is rehearsed naturally whenever the user visits {\em any} site which uses the public cue $\hat{c}$. Lemma~\ref{lemma:ExponentialDistribution} makes use of two key properties of Poisson processes: (1) The natural rehearsal schedule for a cue $\hat{c}$ is itself a Poisson process, and (2) Independent Rehearsals - the probability that a rehearsal constraint is satisfied is independent of previous rehearsal constraints.

\newcommand{\lemmaExponentialDistribution}{Let $S_{\hat{c}} = \{ i ~\vline ~ \hat{c} \in c_i\}$ and let $\lambda_{\hat{c}} = \sum_{i \in S_{\hat{c}}} \lambda_i$ then the probability that the cue $\hat{c}$ is not naturally rehearsed during time interval $\left[a,b\right]$ is $ \exp\paren{-\lambda_{\hat{c}}\paren{b-a}}$.}
\begin{lemma} \label{lemma:ExponentialDistribution}
\lemmaExponentialDistribution
\end{lemma}

\section{Security Model} \label{sec:Security}
In this section we present a game based security model for a password management scheme. The game is played between a user ($\User$) and a resource bounded adversary ($\Adversary$) whose goal is to guess one of the user's passwords. We demonstrate how to select the parameters of the game by estimating the adversary's amortized cost of guessing. Our security definition is in the style of the exact security definitions of Bellare and Rogaway \cite{bellare1996exact}. Previous security metrics (e.g., min-entropy, password strength meters) fail to model the full complexity of the password management problem (see the full version \cite{fullVersion} of this paper for more discussion).  By contrast, we assume that the adversary knows the user's password management scheme and is able to see any public cues. Furthermore, we assume that the adversary has background knowledge (e.g., birth date, hobbies) about the user (formally, the adversary is given $k \in \UserKnowledge$). Many breaches occur because the user falsely assumes that certain information is private (e.g., birth date, hobbies, favorite movie)\cite{breach:palin,secretQuestions:schechter2009s}.

\paragraph{Adversary Attacks.} %\label{subsec:Attacks}
Before introducing our game based security model we consider the attacks that an adversary might mount. We group the adversary attacks into three categories: {\it  Online Attack} --- the adversary knows the user's ID and attempts to guess the password. The adversary will get locked out after $\Strikes$ incorrect guesses (strikes). {\it Offline Attack} --- the adversary learns both the cryptographic hash of the user's password and the hash function and can try many guesses $\Guesses{\$B}$. The adversary is only limited by the resources $B$ that he is willing to invest to crack the user's password. {\it Plaintext Password Leak Attack} --- the adversary directly learns the user's password for an account. Once the adversary recovers the password $p_i$ the account $A_i$ has been compromised. However, a secure password management scheme should prevent the adversary from compromising more accounts.  

\cut{There are many ways that the adversary could recover one of the user's passwords (e.g., phishing, malware, hidden cameras, hidden microphones, shoulder surfing). The adversary could also learn passwords by hacking into sites like RockYou that store passwords in the clear \cite{breach:rockyou}. IEEE and Apple passwords have been stored in the clear due to programmer errors \cite{breach:IEEE,breach:apple}. Once the adversary recovers the password $p_i$ the account $A_i$ is compromised. We do not focus on preventing recovery attacks. Instead we seek to prevent the adversary from compromising more accounts.  }

We model online and offline attacks using a guess-limited oracle. Let $S \subseteq [m]$ be a set of indices, each representing an account. A guess-limited oracle $O_{S,\Guesses{}}$ is a blackbox function with the following behavior: 1) After $\Guesses{}$ queries $O_{S,\Guesses{}}$ stops answering queries. 2) $\forall i \notin S$, $O_{S,\Guesses{}}\paren{i,p} = \bot$ 3) $\forall i \in S$, $O_{S,\Guesses{}}\paren{i,p_i} = 1$ and 4)  $\forall i \in S, p \neq p_i$, $O_{S,\Guesses{}}\paren{i,p} = 0$. Intutively, if the adversary steals the cryptographic password hashes for accounts $\left\{A_i~\vline ~ i \in S \right\}$, then he can execute an offline attack against each of these accounts. We also model an online attack against account $A_i$ with the guess-limited oracle $O_{\{i\}, \Strikes}$ with $\Strikes \ll \Guesses{}$ (e.g., $\Strikes=3$ models a three-strikes policy in which a user is locked out after three incorrect guesses).

\paragraph{Game Based Definition of Security.}
Our cryptographic game proceeds as follows: \\
{\it Setup:} The user $\User$ starts with knowledge $k \in \UserKnowledge$, visitation schedule $\vec{\lambda} \in \mathbb{R}^m$, a random sequence of bits $b \in \{0,1\}^*$ and a rehearsal schedule $R \in \mathcal{R}$. The user runs $\Generator_m\paren{k,b,\vec{\lambda},R}$ to obtain $m$ passwords $p_1,...,p_m$ and public cues $c_1,...,c_m \subseteq \CueSpace$ for accounts $A_1,...,A_m$. The adversary $\Adversary$ is given $k$, $\Generator_m$, $\vec{\lambda}$ and $c_1,...,c_m$. \\
{\it Plaintext Password Leak Attack: } $\Adversary$ adaptively selects a set $S \subseteq [m]$ s.t $\left|S\right| \leq \NumPhish$ and receives $p_i$ for each $i \in S$. \\
{\it Offline Attack: } $\Adversary$ adaptively selects a set $S' \subseteq [m]$ s.t. $\left|S'\right| \leq \NumHashes$, and is given blackbox access to the guess-limited offline oracle $O_{S',\Guesses{}}$. \\
{\it Online Attack: } For each  $i \in [m]- S$, the adversary is given blackbox access to the guess-limited offline oracle $O_{\{i\},\Strikes}$. \\
{\it Winner:}  $\Adversary$ wins by outputting $(j,p)$, where $j \in [m]-S$ and $p = p_j$.

We use $\mathbf{ AdvWins}\paren{k, b,\vec{\lambda}, \Generator_m, \Adversary}$ to denote the event that the adversary wins.

\begin{definition} \label{def:security}
We say that a password management scheme $\Generator_m$ is $(\Guesses{},\delta, m, \Strikes , \NumPhish, \NumHashes)$-secure if for every $k \in \UserKnowledge$ and adversary strategy $\Adversary$ we have
\[ \Pr_{b} \left[  \mathbf{ AdvWins}\paren{k, b,\vec{\lambda}, \Generator_m, \Adversary} \right] \leq \delta \ .\] 
\end{definition} 

\noindent {\bf Discussion: } Observe that the adversary cannot win by outputting the password for an account that he already compromised in a plaintext password leak. For example, suppose that the adversary is able to obtain the plaintext passwords for $\NumPhish = 2$ accounts of his choosing: $p_i$ and $p_j$. While each of these breaches is arguably a success for the adversary the user's password management scheme cannot be blamed for any of these breaches. However, if the adversary can use this information to crack any of the user's other passwords then the password management scheme can be blamed for the additional breaches. For example, if our adversary is also able to use $p_i$ and $p_j$ to crack the cryptographic password hash $h(p_t)$ for another account $A_t$ in at most $\Guesses{}$ guesses then the password management scheme could be blamed for the breach of account $A_t$. Consequently, the adversary would win our game by outputting $(t,p_t)$.  If the password management scheme is  $(\Guesses{},10^{-4}, m, s , 2, 1)$-secure then the probability that the adversary could win is at most $10^{-4}$ --- so there is a very good chance that the adversary will fail to crack $p_t$.

\cut{
\noindent {\bf Example: } Suppose that a password management scheme is  $(\Guesses{\$10^6},10^{-4}, m, s , 3, 1)$-secure and that the adversary is able to recover $\NumPhish = 3$ of the user's plaintext passwords $p_i,p_j,p_k$. Even if the adversary obtains $\NumHashes = 1$ cryptographic password hash $h(p_t)$ and spends $\$10^6$ trying to crack $p_t$ he will fail with probability at least $1-10^{-4}$!}

\paragraph{Economic Upper Bound on $\Guesses{}$.}  Our guessing limit $\Guesses{}$ is based on a model of a resource constrained adversary who has a budget of $\$B$ to crack one of the user's passwords. We use the upper bound $\Guesses{B} = \$B/C_{\Guesses{}}$, where $C_{\Guesses{}} = \$R/{f_H}$ denotes the amortized cost per query (e.g., cost of renting ($\$R$) an hour of computing time on Amazon's cloud \cite{amazonCloudPricing} divided by $f_H$ --- the number of times the cryptographic hash function can be evaluated in an hour.) We experimentally estimate $f_H$ for SHA1, MD5 and BCRYPT\cite{bcrypt} --- more details can be found in the full version \cite{fullVersion} of this paper. Assuming that the BCRYPT password hash function \cite{bcrypt} was used to hash the passwords we get $\Guesses{B} = B\left(5.155\times 10^4\right)$ --- we also consider cryptographic hash functions like SHA1, MD5 in the full version\cite{fullVersion} of this paper.  In our security analysis we focus on the specific value $\Guesses{\$10^6} = 5.155 \times 10^{10}$ --- the number of guesses the adversary can try if he invests $\$10^6$ to crack the user's password.  

\paragraph{Sharing and Security.} In section \ref{sec:usability} we saw that sharing public cues across accounts improves usability by (1) reducing the number of cue-association pairs that the user has to memorize and rehearse, and (2) increasing the rate of natural rehearsals for each cue-association pair. However, conventional security wisdom says that passwords should be chosen independently. Is it possible to share public cues, and satisfy the strong notion of security from definition \ref{def:security}? Theorem \ref{thm:security} demonstrates that public cues can be shared securely provided that the public cues $\{c_1,\ldots,c_m\}$ are a $\paren{\NumBaseCues,\ell,\gamma}$-sharing set family. The proof of theorem \ref{thm:security} can be found in the full version of this paper \cite{fullVersion}. 

\begin{definition} \label{def:GoodSharing} We say that a set family $\mathcal{S} = \{S_1,...,S_m\}$ is $\left(n, \ell, \gamma \right)$-sharing if (1) $\left| \bigcup_{i=1}^m S_i \right| = n$, (2)$\left| S_i \right| = \ell$ for each $S_i \in \mathcal{S}$, and (3) $\left| S_i \cap S_j \right| \leq \gamma$ for each pair $S_i \neq S_j \in \mathcal{S}$.   \end{definition}

 \newcommand{\thmSecurity}{Let $\{c_1,\ldots,c_m\}$ be a $(n,\ell,\gamma)$-sharing set of $m$ public cues produced by the password management scheme $\Generator_m$. If each $a_i \in \AssocSpace$ is chosen uniformly at random then $\Generator_m$ satisfies $(\Guesses{},\delta, m, \Strikes, \NumPhish, \NumHashes)$-security for $\delta \leq \frac{\Guesses{}}{\left| \AssocSpace \right|^{ \ell -\gamma \NumPhish }}$ and any $\NumHashes$.}
\begin{theorem} \label{thm:security}
\thmSecurity
\end{theorem}

{\bf Discussion:} To maintain security it is desirable to have $\ell$ large (so that passwords are strong) and $\gamma$ small (so that passwords remain strong even after an adversary compromises some of the accounts). To maintain usability it is desirable to have $\NumBaseCues$ small (so that the user doesn't have to memorize many cue-association pairs). There is a fundamental trade-off between security and usability because it is difficult to achieve these goals without making $\NumBaseCues$ large.  

For the special case $\NumHashes= 0$ (e.g., the adversary is limited to online attacks) the security guarantees of Theorem \ref{thm:security} can be further improved to $\delta \leq \frac{\Strikes m}{\left| A \right|^{ \ell -\gamma \NumPhish }}$  because the adversary is actually limited to $\Strikes m$ guesses. \cut{For example, $\sharedCues(9,10,11,13)$ is $(\Guesses{},7.6\times 10^{-7}, 90, 2, 3, 0)$-secure and $(\Guesses{},0.015, 90, 3, 3, 0)$-secure --- an adversary who recovered three of the user's plaintext passwords would fail to break any of the user's other passwords except with probability $0.015$.}

\section{Our Construction} \label{sec:PicturesAsCues}

\begin{figure}
        %\centering
        \begin{subfigure}[b]{0.3\textwidth}
			\centering
\includegraphics[scale=0.35]{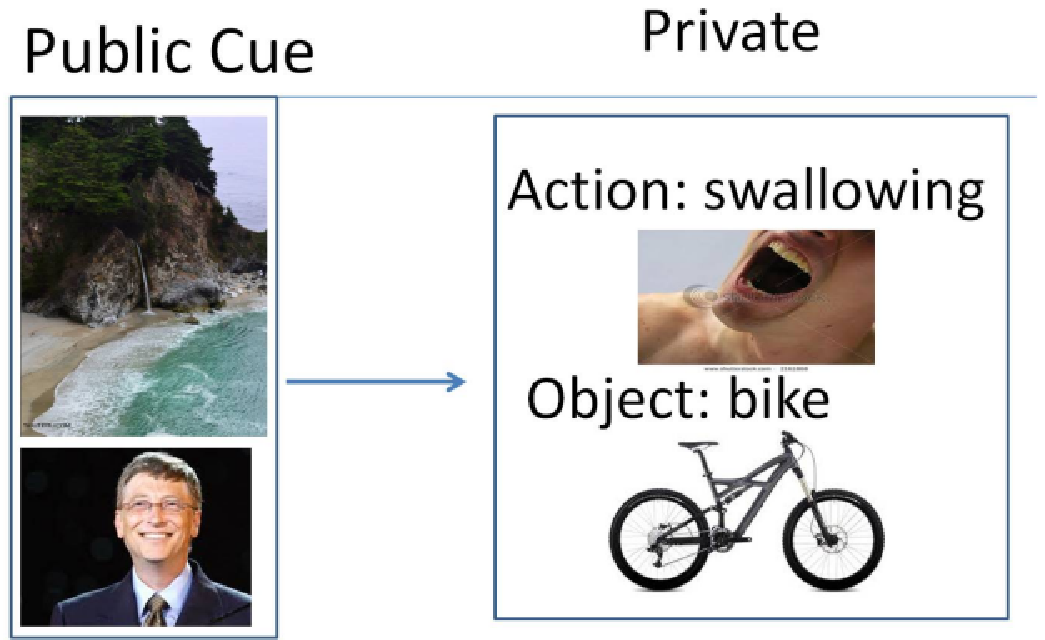}
\caption{PAO Story with Cue} \label{pic:PicturesAsPublicCues}
        \end{subfigure}
        ~ %add desired spacing between images, e. g. ~, \quad, \qquad etc.
        \begin{subfigure}[b]{0.6\textwidth}
                \centering
\includegraphics[scale=0.4]{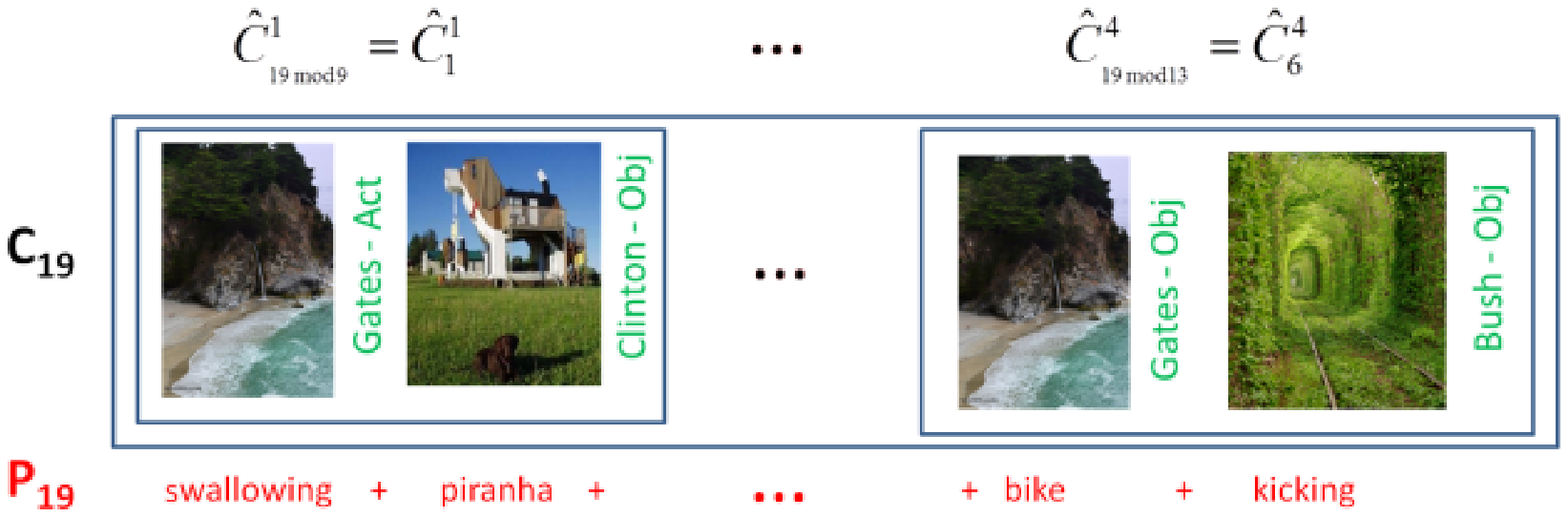}
\caption{ Account $A_{19}$ using \sharedCues~with the $(43,4,1)$-sharing set family $\mathbf{CRT}\paren{90,9,10,11,13}$.}
\label{fig:SharedCues}
        \end{subfigure}
\caption{}
\end{figure}
We present \sharedCues --- a novel password management scheme which balances security and usability considerations. The key idea is to strategically share cues to make sure that each cue is rehearsed frequently while preserving strong security goals. Our construction may be used in conjunction with powerful cue-based mnemonic techniques like memory palaces \cite{memory:memoryPalaceMatteoRicci:spence1985} and person-action-object stories \cite{foer2011moonwalking} to increase $\AssociationStrength{}$ --- the association strength constant.  We use person-action-object stories as a concrete example. 

\cut{
There are several ways that a user might try to ensure that cues are rehearsed sufficiently: 1) The user might commit to rehearse all of his passwords at scheduled times 2) The user might reuse a password (or cue) many times so that it is rehearsed `naturally' each time he logs into an account that uses that password (or cue) 3) The user might employ mnemonic techniques to increase the strength of his associations by minimizing context change and interference. Approach 3 seeks to increase the constant $\AssociationStrength{}$ (association strength).  We strategically share cues to make sure that each cue is rehearsed frequently.  }

\paragraph{Person-Action-Object Stories.}
A random person-action-object (PAO) story for a person (e.g., Bill Gates) consists of  a random action $a \in \Action$  (e.g., swallowing) and a random object $o \in \Object$ (e.g., a bike). While PAO stories follow a very simple syntactic pattern they also tend to be surprising and interesting because the story is often unexpected (e.g., Bill Clinton kissing a piranha, or Michael Jordan torturing a lion). There is good evidence that memorable phrases tend to use uncommon combinations of words in common syntactic patterns \cite{memory:Phrasing}. Each cue $\hat{c} \in \CueSpace$  includes a person (e.g., Bill Gates) as well as a picture. To help the user memorize the story we tell him to imagine the scene taking place inside the picture (see Figure \ref{pic:PicturesAsPublicCues} for an example). We use algorithm \ref{alg:GenStories} to automatically generate random PAO stories. The cue $\hat{c}$ could  be selected either with the user's input (e.g., use the name of a friend and a favorite photograph) or automatically. As long as the cue $\hat{c}$ is fixed before the associated action-object story is selected the cue-association pairs will satisfy the independence condition of Theorem \ref{thm:security}.   %To help the user remember the PAO story we show him the cue $c$ to trigger the story $ao$. 
\cut{
\begin{wrapfigure}{R}{0.4 \textwidth} 
\includegraphics[scale=0.2]{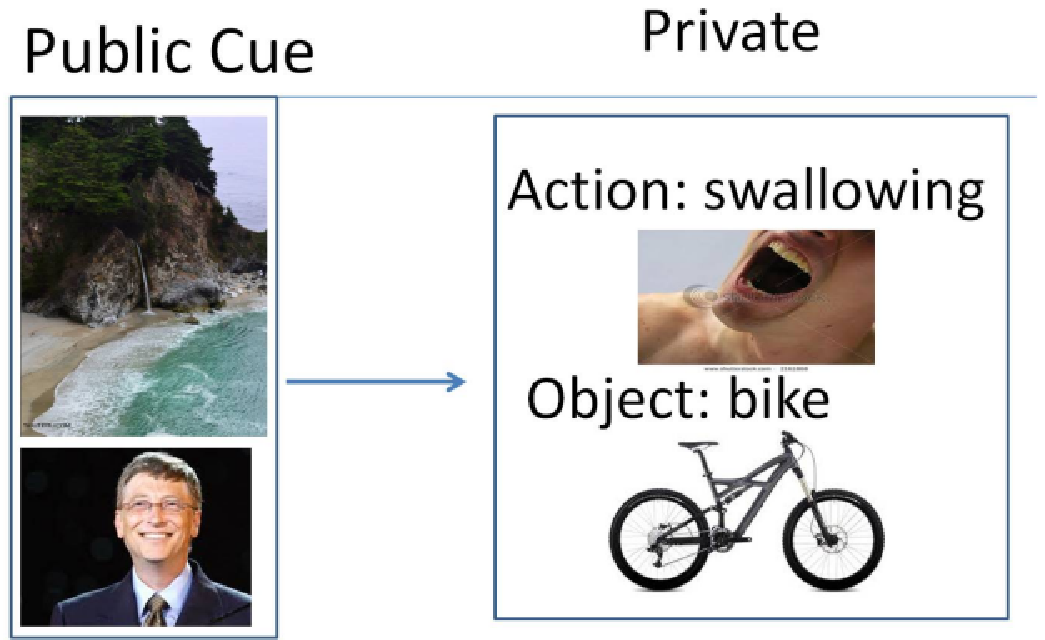}
\caption{PAO Story with Public Cues} \label{pic:PicturesAsPublicCues}
\end{wrapfigure}}

\subsection{Constructing $\paren{n,\ell,\gamma}$-sharing set families}
We use the Chinese Remainder Theorem to construct nearly optimal $\paren{n,\ell,\gamma}$-sharing set families. Our application of the Chinese Remainder Theorem is different from previous applications of the Chinese Remainder Theorem in cryptography (e.g., faster RSA decryption algorithm \cite{ChineseRemainderTheorem}, secret sharing \cite{ChineseRemainderTheorem:SecretSharing}). The inputs $n_1,...,n_\ell$ to algorithm \ref{alg:CRT} should be co-prime so that we can invoke the Chinese Remainder Theorem --- see Figure \ref{fig:SharedCues}  for an example of our construction with $\paren{n_1,n_2,n_3,n_4}=(9,10,11,13)$.
%\begin{figure*}[ttt!]
%\begin{minipage}[t]{2.20in}
\begin{algorithm}[H]
\caption{$\mathbf{CRT}\left(m,n_1,...,n_\ell\right)$}  \label{alg:CRT}
\begin{algorithmic}
\State {\bf Input:} $m$, and $n_1,...,n_\ell$. 
%\Comment{Generate Stories}
\For{$i=1 \to m$}
\State $S_i \gets \emptyset$
\For{$j=1 \to \ell$}
   \State $N_j \gets \sum_{i=1}^{j-1} n_j$
   \State  $S_i \gets S_i \cup \{  \paren{i \mod n_j} + N_j \}$
\EndFor
\EndFor
\Return $\{S_1,\ldots,S_m\}$
\end{algorithmic}
\end{algorithm}
%\end{minipage}
%\hfill 
%\begin{minipage}[t]{3.0in}
\begin{algorithm}[H]
\caption{$\mathbf{CreatePAOStories}$} \label{alg:GenStories}
\begin{algorithmic}
\State {\bf Input:} $n$, random bits b, images $I_1,...,I_n$, and names $P_1,...,P_n$.
 	
%\Comment{Generate Stories}
\For{$i=1 \to n$}
		\State $a_i \stackrel{\$}{\gets} \Action$, $o_i \stackrel{\$}{\gets} \Object$ ~~~\%{\tt Using random bits b} 
   \EndFor
\%{\tt Split PAO stories to optimize usability} 
\For{$i=1 \to n$} \State $\hat{c}_i \gets \left(\left(I_i, P_i,`Act' \right),\left(I_{i+1 \mod n},P_{i+1 \mod n}, `Obj' \right)\right) $ 
\State $\hat{a}_i \gets \paren{a_i, o_{i+1 \mod n}}$ 
\EndFor
\Return $\{\hat{c}_1,\ldots,\hat{c}_n\}, \{\hat{a}_1,\ldots,\hat{a}_n\}$
\end{algorithmic}
\end{algorithm}
% \end{minipage}
% \hfill
%\end{figure*}

Lemma \ref{lemma:Intersection} says that algorithm \ref{alg:CRT} produces a $\paren{n,\ell,\gamma}$-sharing set family of size $m$ as long as certain technical conditions apply (e.g., algorithm \ref{alg:CRT} can be run with any numbers $n_1,...,n_\ell$, but lemma \ref{lemma:Intersection} only applies if the numbers are pairwise co-prime.).

\begin{lemma} \label{lemma:Intersection}
 If the numbers $n_1 < n_2 < \ldots < n_\ell$ are pairwise co-prime and $m \leq \prod_{i=1}^{\gamma+1} n_i$  then algorithm \ref{alg:CRT} returns a $(\sum_{i=1}^\ell n_i,\ell,\gamma)$-sharing set of public cues. 
\end{lemma}
\begin{proof}
Suppose for contradiction that $\left| S_i \bigcap S_k \right| \geq \gamma+1$ for $i < k < m$, then by construction we can find $\gamma+1$ distinct indices $j_1,...,j_{\gamma+1} \in $ such that $i \equiv  k \mod{n_{j_t}}$~~for $1 \leq t \leq \gamma+1$. The Chinese Remainder Theorem states that there is a unique number $x^*$ s.t. (1) $1 \leq x^* < \prod_{t = 1}^{\gamma+1} n_{j_t}$, and (2) $x^* \equiv k \mod{n_{j_t}}$~~for $1\leq t \leq \gamma+1$. However, we have $i < m \leq \prod_{t=1}^{\gamma+1} n_{j_t}$. Hence, $i = x^*$ and by similar reasoning $k = x^*$. Contradiction!
\cut{\begin{align*}
 i \equiv j \mod{n_{k_t}} & \mbox{~~for $1 \leq t \leq u$.}
\end{align*}}

\cut{\begin{align*}
 & \mbox{~~~for each $t \leq u$.}
\end{align*}}

\end{proof}

\noindent{\bf Example:} Suppose that we select pairwise co-prime numbers $n_1 = 9,n_2 = 10, n_3 = 11, n_4 = 13$, then $\mathbf{CRT}\paren{m,n_1,\ldots,n_4}$ generates a $\paren{43,4,1}$-sharing set family of size $m =  n_1\times n_2 = 90$  (i.e. the public cues for two accounts will overlap in at most one common  cue), and for $m \leq n_1 \times n_2 \times n_3 = 990$ we get a $\paren{43,4,2}$-sharing set family. 

Lemma \ref{lemma:Intersection} implies that we can construct a $\paren{n,\ell,\gamma}$-sharing set system of size  $m \geq \Omega\paren{\paren{n/\ell}^{\gamma+1}}$ by selecting each $n_i \approx n/\ell$. 
Theorem \ref{thm:SecurityUpperBound} proves that we can't hope to do much better --- any $\paren{n,\ell,\gamma}$-sharing set system has size $m \leq O\paren{\paren{n/\ell}^{\gamma+1}}$. We refer the interested reader to the full version\cite{fullVersion} of this paper for the proof of Theorem \ref{thm:SecurityUpperBound} and for discussion about additional $(n,\ell,\gamma)$-sharing constructions. 

\newcommand{\thmSecurityUperBound}{Suppose that $\mathcal{S} = \left\{S_1,...,S_m\right\}$ is a $\left(n,\ell,\gamma\right)$-sharing set family of size $m$ then $m \leq {{n \choose {\gamma+1} }} \Big/ {{\ell \choose {\gamma+1} } }$.}
\begin{theorem} \label{thm:SecurityUpperBound}
\thmSecurityUperBound
\end{theorem}

\subsection{Shared Cues}
Our password management scheme ---\sharedCues--- uses a $\paren{n,\ell,\gamma}$-sharing set family of size $m$ (e.g., a set family generated by algorithm \ref{alg:CRT}) as a hardcoded input to output the public cues $c_1,...c_m \subseteq \CueSpace$ and passwords $p_1,...,p_m$ for each account. We use algorithm \ref{alg:GenStories} to generate the underlying cues $\hat{c}_1,\ldots,\hat{c}_n \in \CueSpace$ and their associated PAO stories. The computer is responsible for storing the public cues in persistent memory and the user is responsible for memorizing and rehearsing each cue-association pair $\paren{\hat{c}_i,\hat{a}_i}$. %For notational convenience we use $\sharedCues\left[n_1,\ldots,n_\ell \right]$ to denote the password management scheme when $\mathbf{CRT}\paren{m,n_1,\ldots,n_\ell}$ is used to construct the family of public cues. 
  
We use two additional tricks to improve usability: (1) Algorithm \ref{alg:GenStories} splits each PAO story into two parts so that each cue $\hat{c}$ consists of {\em two} pictures and {\em two} corresponding people with a label (action/object) for each person (see Figure \ref{fig:SharedCues}). A user who sees cue $\hat{c}_i$ will be rehearsing both the $i$'th and the $i+1$'th PAO story, but will only have to enter one action and one object. (2) To optimize usability we use GreedyMap (Algorithm \ref{alg:GreedyMap}) to produce a  permutation $\pi:[m]\rightarrow [m]$ over the public cues --- the goal is to minimize the total number of extra rehearsals by ensuring that each cue is used by a frequently visited account. 

\begin{algorithm}
\caption{$\sharedCues\left[S_1,\ldots,S_m,\right]$ ~ $\Generator_m$}
\begin{algorithmic}
\State {\bf Input:} $k \in \UserKnowledge$, $b$, $\lambda_1,...,\lambda_m$, Rehearsal Schedule $R$.
\State $\{\hat{c}_1,\ldots,\hat{c}_n\}, \{\hat{a}_1,\ldots,\hat{a}_n \} \gets \mathbf{CreatePAOStories\left(n, I_1,...,I_n,P_1,\ldots,P_n\right)}$
%\Comment{Generate Stories}
\For{$i=1 \to m$}
	\State $c_i \gets \left\{ \hat{c}_j ~ \vline~j \in S_i \right\}$, and  $p_i \gets \left\{\hat{a}_j  ~ \vline~j \in S_i \right\} $.
\EndFor
\% {\tt Permute cues }
\State $\pi \gets GreedyMap\left(m, \lambda_1,...,\lambda_m,c_1,\ldots,c_m,R,\AssociationStrength{}\right)$
\State \Return $\left(p_{\pi(1)},c_{\pi(1)}\right),\ldots,\left(p_{\pi(m)},c_{\pi(m)}\right)$
\State {\bf User: } Rehearses the cue-association pairs $\left(\hat{c}_i, \hat{a}_i\right)$ by following the rehearsal schedule $R$.
\State {\bf Computer: } Stores the public cues $c_1,...,c_m$ in persistent memory.
\end{algorithmic}
\label{alg:SharedCues}
\end{algorithm}

Once we have constructed our public cues $c_1,...,c_m \subseteq \CueSpace$ we need to create a mapping $\pi$ between cues and accounts $A_1,...,A_m$. Our goal is to minimize the total number of extra rehearsals that the user has to do to satisfy his rehearsal requirements. Formally, we define the {\bf Min-Rehearsal} problem as follows: \\ 
{\noindent \bf Instance: } Public Cues $c_1,...,c_m \subseteq \CueSpace$, Visitation Schedule $\lambda_1,...,\lambda_m$, a rehearsal schedule $R$ for the underlying cues $\hat{c} \in C$ and a time frame $t$. \\ 
{\noindent \bf Output: } A bijective mapping $\pi:\{1,...,m\}\rightarrow \{1,...,m\}$ mapping account $A_i$ to public cue $S_{\pi(i)} $ which minimizes $E\left[\TotalExtraRehearsals{t}\right]$.\\ Unfortunately, we can show that {\bf Min-Rehearsal} is NP-Hard to even approximate within a constant factor. Our reduction from Set Cover can be found in the full version\cite{fullVersion} of this paper. Instead GreedyMap uses a greedy heuristic to generate a permutation $\pi$.

\newcommand{\HardnessTheorem}{It is NP-Hard to approximate {\bf Min-Rehearsal} within a constant factor.}
\begin{theorem}\label{thm:MinRehearsalHard} \HardnessTheorem
\end{theorem}

%\vspace{-.15in}
\begin{algorithm}
\caption{GreedyMap}
\begin{algorithmic}
\State {\bf Input:}  $m, \lambda_1,...,\lambda_m, c_1,\ldots,c_m$, Rehearsal Schedule $R$ (e.g., CR or ER with parameter $\AssociationStrength{}$). 
\State {\bf Relabel:} Sort $\lambda$'s s.t $\lambda_i \geq \lambda_{i+1}$ for all $i \leq m-1$.
\State {\bf Initialize:} $\pi_0\left(j\right)\gets \bot$ for $j \leq m$, $UsedCues \gets \emptyset$.
\State \%{\tt $\pi_i$ denotes a partial mapping $[i] \rightarrow [m]$,for $j > i$, the mapping is undefined (e.g., $\pi_i\left(j\right) = \bot$). Let $S_k = \left\{\hat{c}~\vline~ \hat{c} \in c_k  \right\}$.}
\For{$i=1 \to m$}
\ForAll{$j \in [m] - UsedCues$}
\State $\Delta_j \gets \displaystyle \sum_{\hat{c} \in S_j} E\left[\ExtraRehearsals{t}{\hat{c}}~\vline \lambda_{\hat{c}} = \lambda_i+ \sum_{j: \hat{c} \in S_{\pi_{i-1}\left(j\right)}} \lambda_j \right] - E\left[\ExtraRehearsals{t}{\hat{c}}~\vline \lambda_{\hat{c}} = \sum_{j: \hat{c} \in S_{\pi_{i-1}\left(j\right)}} \lambda_j\right] $
 \% {\tt $\Delta_j$: expected reduction in total extra rehearsals if we set $\pi_i(i) = j$}
\EndFor
\State $\pi_i\left(i\right) \gets \arg\max_j \Delta_j$, $UsedCues \gets UsedCues\cup \left\{\pi_i\left(i\right)\right\}$
\EndFor 
\Return $\pi^m$
\end{algorithmic}
\label{alg:GreedyMap}
\end{algorithm}
%\vspace{-.1in}
\subsection{Usability and Security Analysis} \label{subsec:usability}
We consider three instantiations of \sharedCues: SC-0, SC-1 and SC-2. SC-0 uses a $\paren{9,4,3}$-sharing family of public cues of size $m=126$ --- constructed by taking all ${9 \choose 4} = 126$ subsets of size $4$. SC-1 uses a $\paren{43,4,1}$-sharing family of public cues of size $m=90$ --- constructed using algorithm \ref{alg:CRT} with $m=90$ and $\paren{n_1,n_2,n_3,n_4} = \paren{9,10,11,13}$. SC-2 uses a $\paren{60,5,1}$-sharing family of public cues of size $m=90$ --- constructed using algorithm \ref{alg:CRT} with $m=90$ and $\paren{n_1,n_2,n_3,n_4,n_5} = \paren{9,10,11,13,17}$.

Our usability results can be found in table \ref{tab:Usability} and our security results can be found in table \ref{tab:Security}. We present our usability results for the very active, typical, occasional and infrequent internet users (see table \ref{tab:userSchedules} for the visitation schedules) under both sufficient rehearsal assumptions CR and ER. Table \ref{tab:Usability} shows the values of $E\left[\TotalExtraRehearsals{365} \right]$ --- computed using the formula from Theorem \ref{thm:ExtraRehearsals} --- for SC-0, SC-1 and SC-2. We used association strength parameter $\AssociationStrength{} = 1$ to evaluate each password management scheme --- though we expect that $\AssociationStrength{}$ will be higher for schemes like \sharedCues~that use strong mnemonic techniques \footnote{We explore the effect of $\AssociationStrength{}$ on $E\left[ \ExtraRehearsals{t}{c} \right]$ in the full version\cite{fullVersion} of this paper.}. 

\begin{table}
\centering
\begin{tabular}{| l | l | l | l || l | l | l | }
\hline 
Assumption & \multicolumn{3}{c|}{CR $(\AssociationStrength{}=1)$ } & \multicolumn{3}{||c|}{ER $(\AssociationStrength{}=1)$} \\
\hline
Schedule/Scheme & SC-0 & SC-1 &  SC-2 & SC-0 & SC-1 & SC-2 \\
\hline
Very Active & $\approx 0$ & $1,309$ & $2,436$ & $\approx 0$ & $3.93$ & $7.54$ \\
\hline
Typical & $\approx 0.42$  & $3,225$ & $5,491$ & $\approx 0$ & $10.89$ & $19.89$ \\
\hline 
Occasional & $\approx 1.28$ &$9,488$ & $6,734$ & $\approx 0$ & $22.07$ & $34.23$ \\
\hline
Infrequent & $\approx 723$ & $13,214$ & $18,764$ & $\approx 2.44$ & $119.77$ & $173.92$  \\
\hline
\end{tabular}
\caption{$E\left[\TotalExtraRehearsals{365}\right]$: Extra Rehearsals over the first year for SC-0,SC-1 and SC-2. }
\label{tab:Usability}
\end{table}

\begin{table}[t]
\centering
\begin{tabular}{| c || c | c | c || c | c | c |}
 \hline 
 Offline Attack? & \multicolumn{3}{|c|}{$\NumHashes = 0$}  &  \multicolumn{3}{|c|}{$\NumHashes > 0$}  \\
\hline
$\paren{n,\ell,\gamma}$-sharing & $\NumPhish = 0$ & $\NumPhish = 1$ & $\NumPhish = 2$ &  $\NumPhish = 0$ & $\NumPhish = 1$ & $\NumPhish = 2$  \\
$\paren{n,4,3}$ (e.g., SC-0) & $2 \times 10^{-15}$  & 0.011 & 1 & $3.5 \times 10^{-7}$ & 1 & 1 \\
$\paren{n,4,1}$ (e.g., SC-1) & $2 \times 10^{-15}$   & $4 \times 10^{-11}$ & $8 \times 10^{-7}$  &  $3.5 \times 10^{-7}$ & 0.007 & 1\\
$\paren{n,5,1}$ (e.g., SC-2) & $1 \times 10^{-19}$  &$2 \times 10^{-15}$ & $4 \times 10^{-11}$ & $1.8\times 10^{-11}$  & $3.5 \times 10^{-7}$ & 0.007    \\
\hline
\end{tabular}
%\vspace{-.1in}
\caption{\sharedCues~$(\Guesses{\$10^6},\delta, m, s , \NumPhish, \NumHashes)$-Security: $\delta$ vs $\NumHashes$ and $\NumPhish$ using a $\paren{n,\ell,\gamma}$-sharing family of $m$ public cues.}
\label{tab:Security}
%\vspace{-0.2in}
\end{table}

Our security guarantees for SC-0,SC-1 and SC-2 are illustrated in Table \ref{tab:Security}. The values were computed using Theorem \ref{thm:security}.  We assume that $\left|\AssocSpace \right| = 140^2$ where $\AssocSpace = \Action \times \Object$ (e.g., their are $140$ distinct actions and objects), and that the adversary is willing to spend at most $\$10^6$ on cracking the user's passwords (e.g., $\Guesses{} = \Guesses{\$10^6}= 5.155 \times 10^{10}$).  The values of $\delta$ in the $\NumHashes = 0$ columns were computed assuming that $m \leq 100$. 

{\bf Discussion:} Comparing tables \ref{tab:Usability} and \ref{tab:UsabilityOld} we see that $\Lifehacker$ is the most usable password management scheme, but SC-0 compares very favorably! Unlike $\Lifehacker$, SC-0 provides provable security guarantees after the adversary phishes one account --- though the guarantees break down if the adversary can also execute an offline attack.  While SC-1 and SC-2 are not as secure as  \StrongRandomPassword~--- the security guarantees from \StrongRandomPassword~do not break down even if the adversary can recover {\em many} of the user's plaintext passwords --- SC-1 and SC-2 are far more usable than \StrongRandomPassword. Furthermore, SC-1 and SC-2 do provide very strong security guarantees (e.g., SC-2 passwords remain secure against offline attacks even after an adversary obtains two plaintext passwords for accounts of his choosing). For the very active, typical and occasional user the number of extra rehearsals required by SC-1 and SC-2 are quite reasonable (e.g., the typical user would need to perform less than one extra rehearsal per month). The usability benefits of SC-1 and SC-2 are less pronounced for the infrequent user --- though the advantage over \StrongRandomPassword~is still significant. 

\cut{
As expected $\Lifehacker$~always requires the fewest number of extra rehearsals --- this is not surprising because there are only $\NumBaseCues = 4$ cues (e.g., three words, one derivation rule) that need to be rehearsed and each cue naturally rehearsed whenever the user logs into any account. By contrast, \StrongRandomPassword~has more rehearsal requirements (e.g., four words for each account $\NumBaseCues=4\times 75 = 300$), and each rehearsal requirement is less likely to be satisfied naturally --- even a very active user has many accounts that he visits infrequently. Our scheme \sharedCues[9,10,11,13]~requires under four extra rehearsals for the very active user, because we only have $\NumBaseCues = 43$ cues and each cue is frequently rehearsed naturally. 

\StrongRandomPassword~ satisfies $(\Guesses{\$10^6},3.2\times 10^{-7}, m, \Strikes, \NumPhish, \NumHashes)$-security for {\em any} values of $\NumPhish$ and $\NumHashes$ --- a very strong security guarantee! By contrast, \Lifehacker~ does not satisfy $(\Guesses{},\delta, m, \Strikes, \NumPhish, 0)$-security whenever $\NumPhish > 0$ because it is easy for the adversary to guess the derivation rule.

}

\section{Discussion and Future Work} \label{sec:discussion}
We conclude by discussing future directions of research.\\ 
{\noindent \bf Sufficient Rehearsal Assumptions:} While there is strong empirical evidence for the Expanding Rehearsal assumption in the memory literature (e.g., \cite{memory:ExpandingRehearsal}), the parameters we use are drawn from prior studies in other domains. It would be useful to conduct user studies to test the Expanding Rehearsal assumption in the password context, and obtain parameter estimates specific to the password setting. We also believe that user feedback from a password management scheme like \sharedCues~could be an invaluable source of data about rehearsal and long term memory retention. 
\\{\noindent\bf Expanding Security over Time: } Most extra rehearsals occur soon after the user memorizes a cue-association pair --- when the rehearsal intervals  are still small. Is it possible to start with a password management scheme with weaker security guaratnees (e.g., SC-0), and increase security over time by having the user memorize additional cue-association pairs as time passes? \\ 
{\noindent \bf Human Computable Passwords: } \sharedCues~ only relies on the human capacity to memorize and retrieve information, and is secure against at most $\NumPhish = \ell/\gamma$ plaintext password leak attacks. Could we improve security (or usability) by having the user perform simple computations to recover his passwords? Hopper and Blum proposed a `human authentication protocol' --- based on the noisy parity problem --- as an alternative to passwords
\cite{hopper2001secure}, but their protocol seems to be too complicated
for humans to execute. Could similar ideas be used to construct a secure human-computation based password management scheme?

\cut{
\noindent {\bf Visitation Schedules:} We considered four types of visitation schedules: the very active user, the typical user, the occasional user and the infrequent user. While \sharedCues~always reduces the burden on the user (compared with \StrongRandomPassword) the reduction is most noticable for a more active user. It would be useful to study user login frequency (e.g., how many users fit the very active profile?). While many sites may track login frequency for each user, we need data at the user level.

\noindent {\bf Other Mnemonic Techniques: } We illustrate the concept of public cues using person-action-object stories with pictures as public cues. However, our technique is not limited to this one mnemonic device. The public cue does not necessarily have to be a picture. It would be interesting to design password management schemes which used musical phrases, or even videos as a public cue. 
}
\cut{
\noindent {\bf Human Authentication Protocols: } One interesting direction is to develop a secure authentication protocol so simple that human's could run the protocol to authenticate themselves in an Orwellian world --- everything the user says or types is monitored so that even biometrics \cite{dodis2004fuzzy} could not be used. 
}

\bibliographystyle{acm}
\bibliography{password}

\appendix

\section{Missing Proofs} \label{appendix:MissingProofs}

\begin{reminderlemma}{\ref{lemma:ExponentialDistribution}}
\lemmaExponentialDistribution
\end{reminderlemma}

\begin{proofof}{Lemma \ref{lemma:ExponentialDistribution}}
Let $N(t) = \left|\left\{\tau^i_k ~\vline~  i \in S_{\hat{c}} \wedge  \tau^i_k \leq t \right\} \right|$ denote the number of times the cue $\hat{c}$ is rehearsed during the interval $[0,t]$. Notice that the rehearsal requirement $[a,b]$ is naturally satisfied if and only if $N(b)-N(a) > 0$. $N(t)$ describes a Poisson arrival process with parameter $\lambda_c =\sum_{i \in S_{\hat{c}}} \lambda_i$ so we can apply standard properties of the Poisson process to get

\[ \Pr\left[N(b)-N(a) = 0 \right] = \exp\paren{-\lambda\paren{b-a}}  \ . \]
\end{proofof}

\begin{remindertheorem}{\ref{thm:ExtraRehearsals}}
\thmExtraRehearsals
\end{remindertheorem}

\begin{proofof}{Theorem \ref{thm:ExtraRehearsals}}
Let $S_{\hat{c}} = \{ i ~\vline ~ \hat{c} \in c_i\}$ and let $V_{a,b}\paren{\hat{c}}$ be the indicator for the event that $\exists i \in S_{\hat{c}}, k \in \mathbb{N}. \tau^i_k \in [a,b]$  (e.g., cue $\hat{c}$ is rehearsed naturally during the time interval $[a,b]$). Then by linearity of expectation
\[E\left[\ExtraRehearsals{t}{\hat{c}}\right] =  \sum_{i=0}^{i_{\hat{c}}*} \paren{1-E\left[V_{t_i,t_{i+1}}\paren{\hat{c}}\right]} \ , \]
where  
\[E\left[1-V_{t_i,t_{i+1}}\paren{\hat{c}}\right] =  \sum_{i=0}^{i_{\hat{c}}*} \exp \left(-\left(\sum_{j:\hat{c} \in c_j} \lambda_j \right)\left(t^{\hat{c}}_{i+1}-t^{\hat{c}}_{i} \right) \right) \ , \]
by Lemma \ref{lemma:ExponentialDistribution}. The result follows immediately from linearity of expectation.
\end{proofof}

\begin{remindertheorem}{\ref{thm:MinRehearsalHard}}
 \HardnessTheorem
\end{remindertheorem}

\begin{proofof}{Theorem \ref{thm:MinRehearsalHard}}
Let $\gamma>0$ be any constant. We prove that it is NP-Hard to even $\gamma$-approximate {\bf Min-Rehearsal}. The reduction is from set cover. \\

{\bf Set Cover Instance:} Sets $S_1,...,S_n$ and universe $U = \bigcup_i S_i$. A set cover is a set $S \subseteq \{1,...,n\}$ such that $\bigcup_{i \in S} S_i = U$. \\
{\bf Question:} Is there a set cover of size $k$?\\

Given a set cover instance, we set $\CueSpace = U$ create public cues $c_1,...,c_m \subseteq \CueSpace$ for each account by setting $c_i = S_i$. We set the following visitation schedule \[\lambda_i = \frac{\ln \left(\gamma \left| U \right| \left(\max_{\hat{c} \in \CueSpace} i_{\hat{c}}^*\right)  \right) }{\min_{j,\hat{c}} \left(t^{\hat{c}}_{j+1}-t^{\hat{c}}_{j}\right)} \ , \] for $i=1,\ldots,k$ and $\lambda_{k+1},...,\lambda_n = 0$. There are two cases: (1) There is a set cover $S =  \{x_1,...,x_k\} \subseteq \{1,...,n\}$ of size $k$. If we assign $\pi(i) = x_i$ for each $i \leq k$ then for each base cue $\hat{c} \in U$ we have \[\lambda_{\hat{c}} = \sum_{i:\hat{c} \in S_i} \lambda_i \geq \lambda_1 \ . \] Applying Theorem \ref{thm:ExtraRehearsals} we get
\begin{eqnarray*}
E\left[\TotalExtraRehearsals{t}\right] &=&= \sum_{\hat{c} \in \CueSpace} \sum_{i=0}^{i_{\hat{c}}*} \exp\left(- \left(t^{\hat{c}}_{i+1}-t^{\hat{c}}_{i}\right) \sum_{i: \hat{c} \in S_{\pi(i)}} \lambda_i \right) \\
&\leq& \left|\CueSpace \right| \left(\max_{\hat{c} \in \CueSpace} i_{\hat{c}}^*\right)  \exp\left(- \left(t^{\hat{c}}_{i+1}-t^{\hat{c}}_{i}\right) \frac{\ln \left(\gamma \left| U \right| \left(\max_{\hat{c} \in \CueSpace} i_{\hat{c}}^*\right)\right)}{\left(\min_{j,\hat{c}} \left(t^{\hat{c}}_{i+1}-t^{\hat{c}}_{i}\right)\right)} \right)  \\
&\leq& \left|U\right| \left(\max_{\hat{c} \in \CueSpace} i_{\hat{c}}^*\right)  \exp\left(-  \ln \left(\gamma \left| U \right| \left(\max_{\hat{c} \in \CueSpace} i_{\hat{c}}^*\right)\right) \right)  \\
&\leq& \left|U \right| \left(\max_{\hat{c} \in \CueSpace} i_{\hat{c}}^*\right)  \frac{1}{ \gamma \left| U \right| \left(\max_{\hat{c} \in \CueSpace} i_{\hat{c}}^*\right)}  \\
&=& \frac{1}{\gamma} \ .
\end{eqnarray*}

(2) If there is no set cover of size $k$. Given a mapping $\pi$ we let $S_\pi = \left\{ i~\vline~\exists j \leq k. \pi\left(j\right) = i\right\}$ be the set of all public cues visited with frequency at least $\lambda_1$. Because $\left|S_\pi\right| = k$,  $S_\pi$ cannot be a set cover and there exists some $\hat{c}_j \in \CueSpace$ which is never visited so no rehearsal requirements are satisfied naturally.
\[E\left[\TotalExtraRehearsals{t}\right]  = \sum_{\hat{c} \in \CueSpace} \sum_{i=0}^{i_{\hat{c}}*} \exp\left(- \left(t^{\hat{c}}_{i+1}-t^{\hat{c}}_{i}\right) \sum_{i: \hat{c} \in S_{\pi(i)}} \lambda_i \right) \geq  \sum_{i=0}^{i_{\hat{c}_j}*} 1 \geq 1\ .\]

\end{proofof}

\begin{remindertheorem}{\ref{thm:security}}
\thmSecurity
\end{remindertheorem}

\begin{proofof}{Theorem \ref{thm:security}}
Recall that $S$ (resp. $S'$) denotes the set of accounts that the adversary selected for plaintext recovery attacks. Let $\paren{k,p_k'}$ denote the adversary's final answer. We can assume that $k \notin S$ because the adversary cannot win by outputting a password he obtained earlier in the game during a plaintext recovery attack. We define 
\[ U_k = c_k - \left\{\hat{c}~\vline~ \exists j \in S.~\hat{c}\in c_j\right\} \ , \]
to be the set of all uncompromised base cues in $c_k$. Observe that

\begin{eqnarray*}
\left|U_k \right| &\geq& \left| c_k\right| - \sum_{j \in S} \left| c_k \bigcap c_j\right| \\
&\geq& \ell - \sum_{j \in S} \gamma \\
&\geq& \ell - \NumPhish \gamma \ ,
\end{eqnarray*}
by definition \ref{def:GoodSharing} of a $\paren{n,\ell,\gamma}$-sharing family of public cues. 

For each, $\hat{c} \in U_k$ the corresponding association $\hat{a}$ was chosen uniformly at random from $\AssocSpace$. We can upper bound $B_\Adversary$ --- the bad event that the adversary $\Adversary$ guesses $(k,p_k)$ in at most $\Guesses{}$ attempts.
\[\Pr\left[ B_\Adversary\right] \leq \frac{\Guesses{}}{\left| \AssocSpace\right|^{\left|U_k\right|}} \leq \frac{\Guesses{}}{\left| \AssocSpace\right|^{\ell - \NumPhish \gamma}} \ . \]
\end{proofof}

\begin{remindertheorem}{\ref{thm:SecurityUpperBound}}
\thmSecurityUperBound
\end{remindertheorem}

\begin{proofof}{Theorem \ref{thm:SecurityUpperBound}}
Let $S \in \mathcal{S}$ be given, and let $T \subseteq S$ be subset of size $\left| T \right| = \gamma+1$. By definition of $\left(n,\ell,\gamma\right)$-sharing we cannot have $T \subseteq S'$ for any other set $S' \in \mathcal{S}-S$. In total there are ${n \choose {\gamma+1} }$ subsets of $[n]$ of size $\gamma+1$ and each $S \in \mathcal{S}$ contains ${\ell \choose {\gamma+1} }$ of them.  The result follows from the pigeonhole principle.
\end{proofof}

\section{Varying the Association Strength Constant} 
\label{appendix:AssociationStrength}

In tables \ref{tab:UsabilityOld} and \ref{tab:Usability} we used the same association strength constant for each scheme $\AssociationStrength{} = 1$ --- though we expect that $\AssociationStrength{}$ will be higher for schemes like \sharedCues~that use strong mnemonic techniques. We explore the effect of $\AssociationStrength{}$ on $E\left[ \ExtraRehearsals{t}{c} \right]$ under various values of the nataural rehearsal rate $\lambda$. Table \ref{tab:ErAssoc} shows the values $E\left[ \ExtraRehearsals{t}{c} \right]$ under the expanding rehearsal assumption for $\AssociationStrength{} \in \{0.1.0.5,1,2\}$. We  consider the following natural rehearsal rates: $\lambda = 1$ (e.g., naturally rehearsed daily),  $\lambda = 3$,  $\lambda = 7$ (e.g., naturally rehearsed weekly),  $\lambda = 31$ (e.g., naturally rehearsed monthly). 

\begin{table}
\centering
\begin{tabular*}{4.65in}[h]{|c|c|c|c|c|c|}
\hline
$\lambda$ (visits/days) & $2$ & $1$ & $\frac{1}{3}$ & $\frac{1}{7}$ & $\frac{1}{31}$ \\
\hline
$\AssociationStrength = 0.1$ & 0.686669 & 2.42166 & 5.7746 & 7.43555 & 8.61931 \\
$\AssociationStrength = 0.5$ & 0.216598 & 0.827594 &   2.75627 & 4.73269 & 7.54973 \\
 $\AssociationStrength = 1$& 0.153986 & 0.521866 & 1.56788 & 2.61413 &   4.65353 \\
$\AssociationStrength = 2$ & 0.135671 & 0.386195 & 0.984956 & 1.5334 & 2.57117 \\
\hline
\end{tabular*}
\caption{Expanding Rehearsal Assumption: $\ExtraRehearsals{365}{c}$ vs. $\lambda_c$ and $\AssociationStrength{}$}
\label{tab:ErAssoc}
\end{table}

Table \ref{tab:CrAssoc} shows the values $E\left[ \ExtraRehearsals{t}{c} \right]$ under the constant rehearsal assumption for $\AssociationStrength{} \in \{1,3,7,31\}$ (e.g., if $\AssociationStrength{} = 7$ then the cue must be rehearsed every week).

\begin{table}
\centering 
\begin{tabular*}{4.6in}[h]{|c|c|c|c|c|c|}
\hline
$\lambda$ (visits/days) & $2$ & $1$ & $\frac{1}{3}$ & $\frac{1}{7}$ & $\frac{1}{31}$ \\
\hline
$\AssociationStrength{} = 1$ &49.5327& 134.644& 262.25& 317.277& 354.382\\
$\AssociationStrength{} = 3$ & 0.3024 & 6.074 & 44.8813 & 79.4756 & 110.747 \\
 $\AssociationStrength{} = 7$& 0.0000 & 0.0483297 & 5.13951 & 19.4976 & 42.2872 \\
$\AssociationStrength{} = 31$ & 0.000 & 0.0000 & 0.0004 & 0.1432 & 4.4146 \\
\hline
\end{tabular*}
\caption{Constant Rehearsal Assumption: $\ExtraRehearsals{365}{c}$ vs. $\lambda_c$ and $\AssociationStrength{}$}
\label{tab:CrAssoc}
\end{table}

\section{Baseline Password Management Schemes} \label{subsec:EvaluateSecurity}
In this section we formalize our baseline password management schemes: \ReuseWeakPassword~(Algorithm \ref{alg:ReuseWeak}), \ReuseStrongPassword~(Algorithm \ref{alg:ReuseStrong}),\Lifehacker~(Algorithm \ref{alg:Lifehacker}) and \StrongRandomPassword~(Algorithm \ref{alg:StrongRandomIndependent}). The first three schemes (\ReuseWeakPassword,\ReuseStrongPassword,\Lifehacker) are easy to use, but only satisfy weak security guarantees. \StrongRandomPassword~ provides very strong security guarantees, but is highly difficult to use.

Vague instructions and strategies do not constitute a password management scheme because it is unclear what the resulting distribution over $\PasswordSpace$ looks like. When given such vague instructions (e.g., ``pick a random sentence and use the first letter of each word'') people tend to behave predictably (e.g., picking a popular phrase from a movie or book). For example, when people are required to add special symbols to their passwords they tend to use a small set of random symbols and add them in predictable places (e.g., end of the password) \cite{usability:compositionPolicies}. Most password advice provides only vague instructions. However, many of these vague strategies can be tweaked to yield formal password management schemes. \ReuseWeakPassword, \ReuseStrongPassword, and \Lifehacker~are formalizations of popular password management strategies.

Each of these password management schemes ignores the visitation schedule $\lambda_1,...,\lambda_m$. None of the schemes use cues explicitly. However, the user always has an implicitly cue when he tries to login. For example, the implicit cue in \ReuseWeakPassword~might be ``that word that I always use as my password." We use four implicit cues for \ReuseStrongPassword~to represent the use of four separate words (chunks \cite{memory:chunks:miller1956}). These implicit cues are shared across all accounts --- a user rehearses the implicit association(s) when he logs into any of his accounts.

\begin{algorithm}
\caption{\ReuseWeakPassword~$\Generator_m$}
\begin{algorithmic}
\State {\bf Input:} Background knowledge $k \in \UserKnowledge$ about the user. Random bits $b$, $\lambda_1,...,\lambda_m$. 
\State {\bf Random Word: } $w \stackrel{\$}{\gets} D_{20,000}$.
\Comment{Select $w$ uniformly at random from a dictionary of 20,000 words.}
\For{$i=1 \to m$}
	\State $p_i \gets w$
   \State $c_i \gets \{`word'\}$
\EndFor
\Return $\left(p_1,c_1\right),...,\left(p_m, c_m\right)$
\State {\bf User: } Memorizes and rehearses the cue-association pairs $\left(`word',p_i\right)$ for each account $A_i$ by following the rehearsal schedule (e.g., CR or ER).
\end{algorithmic}
\label{alg:ReuseWeak}
\end{algorithm}

\begin{algorithm}
\caption{\ReuseStrongPassword~$\Generator_m$}
\begin{algorithmic}
\State {\bf Input:} Background knowledge $k \in \UserKnowledge$ about the user. Random bits $b$, $\lambda_1,...,\lambda_m$. 
\For{$i=1 \to 4$}
	\State {\bf Random Word: } $w_i  \stackrel{\$}{\gets} D_{20,000}$. 
\EndFor
\For{$i=1 \to m$}
	\State $p_i \gets w_1w_2w_3w_4$
   \State $c_i \gets  \left\{\left(`Word',j\right)~\vline~j \in [4]\right\}$
\EndFor
\Return $\left(p_1,c_1\right),...,\left(p_m, c_m\right)$
\State {\bf User: } Memorizes and rehearses the cue-association pairs $\left(\left(`Word',j \right),w_j\right)$ for each $j \in [4]$ by following the rehearsal schedule (e.g., CR or ER).
\end{algorithmic}
\label{alg:ReuseStrong}
\end{algorithm}

\Lifehacker~uses a derivation rule to get a different password for each account. There is no explicit cue to help the user remember the derivation rule, but the implicit cue (e.g., ``that derivation rule I {\em always} use when I make passwords") is shared across every account --- the user rehearses the derivation rule every time he logs into one of his accounts. There are four base cues --- three for the words, one for the derivation rule.

\begin{algorithm}
\caption{\Lifehacker~$\Generator_m$}
\begin{algorithmic}
\State {\bf Input:} Background knowledge $k \in \UserKnowledge$ about the user. Random bits $b$, $\lambda_1,...,\lambda_m$. 
\For{$i=1 \to 3$}
	\State {\bf Random Word: } $w_i \stackrel{\$}{\gets} D_{20,000}$.
\EndFor
\State {\bf Derivation Rule:} $d  \stackrel{\$}{\gets} DerivRules$.
\Comment{$DerivRules$ is a set of $50$ simple derivation rules to map the name of a site $A_i$ to a string $d\left(A_i\right)$ (e.g., use the first three consonants of $A_i$).}
\For{$i=1 \to m$}
	\State $p_i \gets w_1w_2w_3d\left(A_i\right)$
   \State $c_i \gets  \left\{\left(`Word',j\right)~\vline~j \in [3]\right\}\cup \{`Rule'\}$
\EndFor
\Return $\left(p_1,c_1\right),...,\left(p_m, c_m\right)$
\State {\bf User: } Memorizes and rehearses the cue-association pairs $\left(\left(`Word',j\right),w_j\right)$ for each $j \in [3]$ and $(`Rule',d)$ by following the rehearsal schedule (e.g., CR or ER).
\end{algorithmic}
\label{alg:Lifehacker}
\end{algorithm}

\StrongRandomPassword~also uses implicit cues (e.g., the account name $A_i$), which are {\em not} shared across accounts so the only way to naturally rehearse the association $\left(A_i, p_i\right)$ is to visit account $A_i$. 

\begin{algorithm}
\caption{\StrongRandomPassword~$\Generator_m$}
\begin{algorithmic}
\State {\bf Input:} Background knowledge $k \in \UserKnowledge$ about the user. Random bits $b$, $\lambda_1,...,\lambda_m$. 

\For{$i=1 \to m$}
\For{$j=1 \to 4$}
	\State {\bf Random Word: } $w^i_j \stackrel{\$}{\gets} D_{20,000}$.
\EndFor
	\State $p_i \gets w^i_1w^i_2w^i_3w^i_4$
   \State $c_i \gets \left\{\left(A_i,j\right)~\vline~j \in [4]\right\}$
\EndFor
\Return $\left(p_1,c_1\right),...,\left(p_m, c_m\right)$
\State {\bf User: } Memorizes and rehearses the association $\left(\left(A_i,j\right),w^i_j\right)$ for each account $A_i$ and $j \in [4]$ by following the rehearsal schedule (e.g., CR or ER).
\end{algorithmic}
\label{alg:StrongRandomIndependent}
\end{algorithm}

\subsection{Security Of Baseline Password Management Schemes}
\ReuseWeakPassword~is not $(\Guesses{\$1},\delta,m,s,0,1)$-secure for any $\delta < 1$ --- an adversary who is only willing to spend $\$1$ on password cracking will still be able to crack the user's passwords! While \ReuseWeakPassword~does provide some security guarantees against online attacks they are not very strong. For example, \ReuseWeakPassword~is not even $\left(\Guesses{\$1},.01 , 100,3,0,0\right)$-secure because an adversary who executes an online attack can succeed in breaking into at least one of the user's 100 accounts with probability at least $.01$ --- even if all accounts implement a 3-strike limit. If the adversary recovers any of the user's passwords ($\NumPhish > 0$) then all security guarantees break down.

\ReuseStrongPassword~is slightly more secure. It satisfies $(\Guesses{\$10^6},3.222\times 10^{-7},m,s,0,m)$-security meaning that with high probability the adversary who has not been able to recover any of the user's passwords will not even be able to mount a successful offline attack against against the user. However, \ReuseStrongPassword~is not $\paren{\Guesses{},\delta, m, s, 1, 0}$-secure --- if the adversary is able to recover just one password $p_i$ for any account $A_i$ then the adversary will be able to compromise all of the user's accounts.

\Lifehacker~is supposed to limit the damage of a recovery attack by using a derived string at the end of each password. However, in our security model the adversary knows that the user used \Lifehacker~to generate his passwords. The original article \cite{guideline:lifehacker} instructs users to pick a simple derivation rule (e.g., ``user ther first three consonants in the site name"). Because this instruction is vague we assume that there are a set of $50$ derivation rules and that one is selected at random. If the adversary sees a password $p_i = w_1w_2w_3d\left(A_i\right)$ for account $A_i$ then he can immediately infer the base password $b = w_1w_2w_3$, and the adversary needs at most $50$ guesses to discover one of the user's passwords\footnote{In fact the adversary most likely needs far fewer guesses. He can immediately eliminate any derivation rule $\hat{d}$ s.t. $\hat{d}\left(A_i\right) \neq d\left(A_i\right)$. Most likely this will include almost all derivation rules besides the correct one.} --- so if $(m-1)\Strikes \geq 50$ then \Lifehacker~is not $\left(\Guesses{}, \delta, m ,\Strikes, 1, 0\right)$-secure for any values of $\delta,\Guesses{}$. \Lifehacker~is $(\Guesses{\$10^6},1.29\times 10^{-4},m,s,0,m)$-secure --- it defends against offline and online attacks in the absence of recovery attacks. 

\StrongRandomPassword~is highly secure! It satisfies $(\Guesses{\$10^6},3.222\times 10^{-7},m,\Strikes,\alpha,m)$-security for any $\alpha \leq m$. This means that even after the adversary learns many of the user's passwords he will fail to crack any other password with high probability. Unfortuanately, \StrongRandomPassword is very difficult to use.

\subsection{Usability of Baseline Schemes}
Usability results for \Lifehacker~and \StrongRandomPassword~can be found in table \ref{tab:UsabilityOld} of the paper. We evaluate usability using the formula from Theorem \ref{thm:ExtraRehearsals}. We present our results for the Very Active, Typical, Occasional and Infrequent users under both sufficient rehearsal assumptions CR and ER --- with association strength $\AssociationStrength{} = 1$. The usability results for $\ReuseStrongPassword$~are identical to $\Lifehacker$, because they have the same number of cues and each cue is rehearsed anytime the user visits any account $A_i$. Similarly, the usability results for $\ReuseWeakPassword$~are better by a factor of $4$ (e.g., because there is only one cue-association pair  to rehearse and the natural rehearsal rates are identical). 

\subsection{Sources of Randomness} \label{subsec:HumanEntropy}
Popular password advice tends to be informal --- the user is instructed to select a character/number/digit/word, but is not told how to do this. Certainly one reason why people do not select random passwords is because they worry about forgetting their password  \cite{kruger2008empirical}. However, even if the user is told to select a the character uniformly at random it is still impossible to make any formal security guarantees without understanding the entropy of a humanly generated random sequence. We have difficulty consciously generating a random sequence of numbers even when they are not trying to construct a memorable sequence \cite{wagenaar1972generation} \cite{seventeenMostRandom} \cite{humanRandom:figurska2008humans}. 

This does not rule out the possibility that human generated random sequence could provide a weak source of entropy \cite{halprin2010games} --- which could be used to extract a truly random sequence with computer assistance \cite{shaltiel2004recent,dodis2004fuzzy}. We envision a computer program being used to generate random words from a dictionary or random stories (e.g., Person-Action-Object stories) for the user to memorize. The source of randomness could come from the computer itself or it could be extracted from a human source (e.g., a user randomly typing on the keyboard).

  \cut{ 

\noindent {\bf Security of Current Schemes:} We formally evaluate the security of \ReuseWeakPassword, \ReuseStrongPassword, \Lifehacker and \StrongRandomPassword in the appendix (see section \ref{subsec:EvaluateSecurity}). The upshot is that \ReuseWeakPassword, \ReuseStrongPassword, \Lifehacker are not secure. \ReuseWeakPassword is not secure against offline attacks - even if the adversary is only willing to invest $\$1$ to crack the password. \ReuseStrongPassword and \Lifehacker are slightly more secure, but once the adversary compromises just one of the user's passwords all other accounts are vulnerable. \StrongRandomPassword is highly secure! It satisfies $(Q_{\$10^6},3.222\times 10^{-7},m,s,\alpha,m)$-security for any $\alpha < m$, meaning that an adversary who is willing to invest $\$10^6$ will fail to crack any of the user's passwords, and this guarantee hold even after the adversary sees as many example passwords as he wants. }

\section{$\paren{n,\ell,\gamma}$-sharing sets} \label{apdx:nlgGoodSets}
In this section we discuss additional $\paren{n,\ell,\gamma}$-sharing set family constructions. Theorem \ref{thm:CRTImprovement} demonstrates how our Chinese Remainder Theorem construction can be improved slightly. For example, we can get a  $\paren{43,4,1}$-sharing set family of size $m=110$ with the additional optimizations from Theorem \ref{thm:CRTImprovement} --- compared with $m=90$ without the optimizations. We also use a greedy algorithm to construct $\paren{n,\ell,\gamma}$-sharing set families for smaller values of $n$. Our results our summarized in table \ref{tbl:nlgGoodConstructions} --- we also include the theoretical upper bound from theorem \ref{thm:SecurityUpperBound} for comparison.
 
Our notion of $(n,\ell,\gamma)$-sharing set families (definition \ref{def:GoodSharing}) is equivalent to Nisan and Widgerson's definition of a $(k,m)$-design \cite{setSharing}. Nisan and Widgerson provided several constructions of $(k,m)$-designs. For example, one of their constructions implies that their is a $\paren{n,\ell,\gamma}$-sharing set family of size $m=2^h$ for $n = h^{2c},\ell = h^c,$ and $\gamma = h$, where $h$ is any power of $2$ and $c>1$ is a constant. While this construction is useful for building pseudorandom bit generators, it is not especially helpful in the password context because $\ell$ should be a small constant. Even if the user has at most $m = 16$ accounts we would still need at least $\ell = 16$ public cues per account ($c =2$, $h=4$).

\begin{table}
\centering

\begin{tabular}{|c|c|c|c|}
\hline 
$\paren{n,\ell,\gamma}$-sharing & m---Lower Bound & m---Upper Bound (Thm \ref{thm:SecurityUpperBound}) & Comment \\
\hline
$\paren{n,\ell,\ell-1}$ & ${n \choose \ell}$ & ${n \choose \ell}$ & Claim \ref{claim:lminus1good} \\
\hline 
$\paren{9,4,3}$ & 126 & 126 & Greedy Construction (Alg \ref{alg:GreedyConstruction}) \\
\hline 
$\paren{16,4,1}$ & 16 & 20 & Greedy Construction (Alg \ref{alg:GreedyConstruction}) \\
\hline 
$\paren{20,6,2}$ & 40 & 57 & Greedy Construction (Alg \ref{alg:GreedyConstruction}) \\
\hline
$\paren{25,6,2}$ & 77 & 153 & Greedy Construction (Alg \ref{alg:GreedyConstruction}) \\
\hline 
$\paren{18,6,3}$ & 88 & 204 & Greedy Construction (Alg \ref{alg:GreedyConstruction}) \\
\hline
$\paren{19,6,3}$ & 118 & 258 & Greedy Construction (Alg \ref{alg:GreedyConstruction}) \\
\hline
$\paren{30,9,3}$ & 36 & 217& Greedy Construction (Alg \ref{alg:GreedyConstruction}) \\
\hline 
$\paren{40,8,2}$ & 52 & 176 & Greedy Construction (Alg \ref{alg:GreedyConstruction}) \\
\hline 
$\paren{43,4,1}$ & 110 & 150 & Theorem \ref{thm:CRTImprovement} \\
\hline 
\end{tabular}
\caption{$\paren{n,\ell,\gamma}$-sharing set family constructions}  \label{tbl:nlgGoodConstructions}
\end{table}

\begin{algorithm}
\caption{Greedy Construction}
\begin{algorithmic}
\State {\bf Input:}  $n, \ell, \gamma$
\State {\bf All Subsets:} $\mathcal{S}' \gets \left\{ S \subseteq [n] ~\vline ~ \left| S \right| = \ell \right\}$
\State {\bf Candidates: }  $\mathcal{S} \gets \emptyset$
\ForAll{$S \in \mathcal{S}'$}
\State $okToAdd \gets True$
\ForAll{$T \in \mathcal{S}$}
\If{$\left| T \bigcap S\right| > \gamma$} \State $okToAdd \gets False$ \EndIf
\EndFor
\If{$okToAdd$} \State $\mathcal{S} \gets \mathcal{S} \cup \{S\}$ \EndIf
\EndFor
\Return $\mathcal{S}$
\end{algorithmic}
\label{alg:GreedyConstruction}
\end{algorithm}

\begin{theorem} \label{thm:CRTImprovement}
Suppose that $n_1 < \ldots < n_\ell$ are pairwise co-prime and that for each $1 \leq i \leq \ell$ there is a $\paren{n_i,\ell,\gamma}$-sharing set system of size $m_i$. Then there is a  $\paren{\sum_{i=1}^\ell n_i,\ell,\gamma}$-sharing set system of size $m = \prod_{i=1}^\gamma n_i + \sum_{i=1}^\ell m_i$.
\end{theorem}
\begin{proof}
We can use algorithm \ref{alg:CRT} to construct a $\paren{n,\ell,\gamma}$-sharing set family $\mathcal{S}_0$ of size $m' = \prod_{i=1}^\gamma n_i $. Let $T_1 = \left\{k ~\vline k < n_1 \right\}$ and for each $i>1$ let $T_i = \left\{k + \sum_{j=1}^{i-1} n_j~\vline k < n_i \right\}$. By construction of $\mathcal{S}_0$ it follows that for each $S \in \mathcal{S}_0$ and each $1 \leq i \leq \ell$ we have $\left| S \bigcap T_i \right| = 1$. By assumption, for each $i \geq 1$ there is a $\paren{n,\ell,\gamma}$-sharing family of subsets of $T_i$ of size $m_i$ --- denoted $\mathcal{S}_i$. For each pair $S' \in \mathcal{S}_i$, and $S \in \mathcal{S}_0$ we have

\[ \left|S \bigcap S'  \right| \leq   \left|S \bigcap T_i  \right| \leq 1  \ , \]
and for each pair $S' \in \mathcal{S}_i$, and $S \in \mathcal{S}_i$ $(S \neq S')$ 
\[ \left|S \bigcap S'  \right| \leq \gamma  \ , \]
because $\mathcal{S}_i$ is $(n_i,\ell, \gamma)$-sharing.
Finally,  for each pair $S' \in \mathcal{S}_i$, and $S \in \mathcal{S}_j$ ($j\neq i$) we have
\[ \left|S \bigcap S'  \right| \leq   \left|S \bigcap T_i  \right| \leq 0  \ . \]
Therefore, 
\[ \mathcal{S} = \bigcup_{i=0}^\ell \mathcal{S}_i \ , \]
is a $\paren{\sum_{i=1}^\ell n_i,\ell,\gamma}$-sharing set system of size $m = \prod_{i=1}^\gamma n_i + \sum_{i=1}^\ell m_i$.
\end{proof}

\begin{claim} \label{claim:lminus1good}
For any $0 < \ell \leq n$ there is a $\paren{n,\ell,\ell-1}$-sharing set family of size $m = {n \choose \ell}$, and there is no  $\paren{n,\ell,\ell-1}$-sharing set family of size $m' > m$.
\end{claim}
\begin{proof}
 It is easy to verify that \[\mathcal{S} = \left\{ S \subseteq [n]~\vline ~\left| S \right| = \ell \right\} \ ,\]
the set of all subsets of size $\ell$, is a $\paren{n,\ell,\ell-1}$-sharing  set family of size $m = {n \choose \ell}$. Optimality follows immediately by setting $\gamma = \ell-1$ in Theorem \ref{thm:SecurityUpperBound}.
\end{proof}

\section{Other Measures of Password Strength} \label{apdx:OtherSecurity}
In this section we discuss other security metrics (e.g., entropy, minimum entropy, password strength meters, $\alpha$-guesswork) and their relationship to our security model. 

Our security model is fundamentally different from metrics like guessing entropy (e.g., How many guesses does an adversary need to guess all of passwords in a dataset \cite{massey1994guessing}?) and partial guessing entropy (e.g., How many guesses does the adversary need to crack $\alpha$-fraction of the passwords in a dataset \cite{pliam2000incomparability,bonneau2012science}? How many passwords can the adversary break with $\beta$ guesses per account \cite{boztas1999entropies}?), which take the perspective of a system administrator who is trying to protect many users with password protected accounts on his server. For example, a system administrator who wants to evaluate the security effects of a a new password composition policy may be interested in knowing what fraction of user accounts are vulnerable to offline attacks. By contrast, our security model takes the perspective of the user who has many different password protected accounts. This user wants to evaluate the security of various password management schemes that he could choose to adopt. 

Our threat model is also strictly stronger than the threat models behind metrics like $\alpha$-guesswork because we consider targeted adversary attacks from an adversary who may have already compromised some of the user's accounts. 

Password strength meters can provide useful feedback to a user (e.g., they rule out some insecure password management schemes). However, password strength meters are insufficient for our setting for several reasons: (1) They fail to rule out some weak passwords, and (2) They cannot take correlations between a user's passwords (e.g., Is the user reusing the same password?) into account. (3) They do not model the adversaries background knowledge about the user (e.g., Does the adversary know the user's birth date or favorite hobbies?). Entropy is bad measure of security for the same reasons. While minimum entropy fixes some of these problems, minimum entropy still does not address problem 2 --- minimum entropy does not deal with correlated user passwords.

\subsection{Password Strength Meters}
\begin{figure}[h] 
\includegraphics[scale=1]{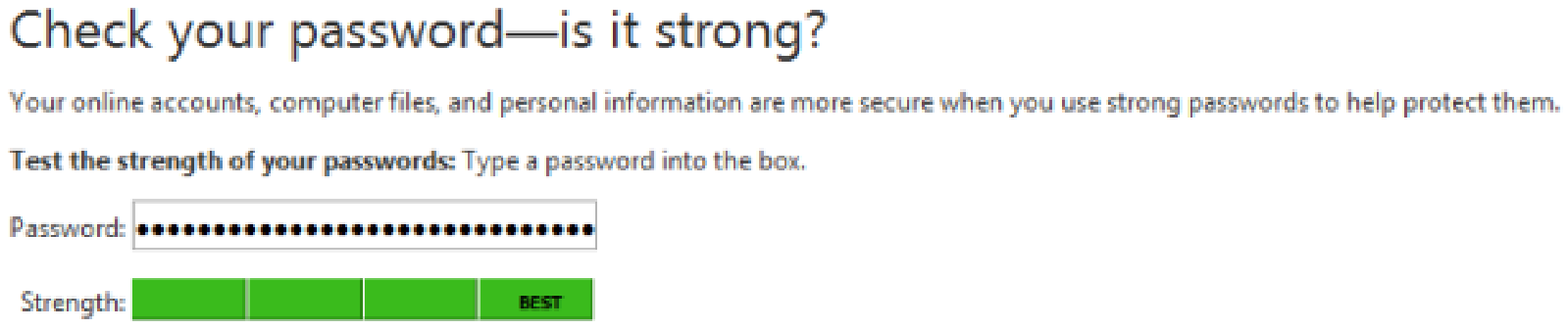}
\caption{mmmmmmmmmmmmmmmmmmmmmmmmmmmm: sounds delicious, but is it really a  strong password?}
\label{picture:PasswordMeter}
\end{figure}

Password strength meters use simple heuristics (e.g., length, character set) to estimate the entropy of a password. A password strength meter can provide useful feedback to the user by warning the user when he picks passwords that are easy to guess. However, password strength meters can also give users a false sense of confidence (e.g.,  `mmmmmmmmmmmmmmmmmmmmmmmmmmmm' is clearly predictable, but is ranked `Best' by some meters \cite{microsoftPasswordStrength} ---  see figure \ref{picture:PasswordMeter} \cite{microsoftPasswordStrength}). A password like {\em Mm1!Mm1!Mm1!Mm1!Mm1!Mm1!} would be rated as very secure by almost any password strength meter because it is long, it uses upper case and lower case letters and it includes a special symbol (!). However, the password is based on a very simple repeated pattern and has low entropy (e.g., it could be compressed easily). A password strength meter cannot guarantee that a password is secure because (1) It does not know whether or not the user has already used this password (or a very similar password) somewhere else (2) It does not know if the user is basing his password on personal knowledge (e.g., wife's birthday) (3) It does not know what background knowledge the adversary might have about the user (e.g., does the adversary know the user's wife's birthday).

\subsection{Entropy} \label{subsec:Entropy}
Entropy \cite{shannon1959mathematical} can be used to measure the average number of guesses an adversary would need to guess a password chosen at random from a distribution $D$ over passwords
\[H\left(D \right) = \sum_x \Pr\left[ x~ \vline~ D\right] \log_2 \left(\frac{1}{\Pr\left[ x \vline~ D\right]} \right) \ . \]   
While entropy has been a commonly used information theoretic measure of password strength \cite{XKCDhorsebatterystaplecorrect,usability:compositionPolicies}, it is not always a good indicator of password strength \cite{massey1994guessing}. For example, consider the following distributions over binary passwords $D_1$ and $D_2$:
\begin{equation*} D_1\left(n\right) = 
\begin{cases} 
1^{n-1} & \text{ with probability $1/2$,} \\
x \in \{0,1\}^{2n-2} & \text{ with probability $2^{-2n+1}$.}
\end{cases} 
\end{equation*}
\[D_2\left(n\right) = x\in\{0,1\}^n~~~\text{ with probability $2^{-n}$} \ .\]
While there is no difference in the entropy of both generators
\[\Entropy\left(D_1\left( n\right) \right) = \frac{1}{2} \log_2 \left(\frac{1}{1/2}\right) +  \sum_{x} 2^{-2n+1} \log_2 \left(2^{2n-1}\right) = \frac{1}{2} + \frac{2n-1}{2} = n  = H\left(D_2\left( n\right)  \right) \ , \]
$D_1$ and $D_2$ are by no means equivalent from a security standpoint! After just one guess an adversary can successfully recover the password generated by $D_1$ with probability $\geq \frac{1}{2}$! By contrast an adversary would need at least $2^{n-1}$ guesses to recover the password generated by $D_2$ with probability $\geq \frac{1}{2}$.
\subsection{Minimum Entropy} \label{subsec:MinEntropy}
If we instead consider the minimum entropy
\[\MinEntropy\left(G\right) = \min_x \log_2 \left(\frac{1}{\Pr\left[x \vline ~ G \right]} \right) \ , \]
 of both generators we get a different story.

\[\MinEntropy\left(D_1\left( n\right)\right) =   \log_2 \left(\frac{1}{1/2}\right) = 1 \ll \MinEntropy\left(D_2\left( n\right)\right) =  \log_2 \left( 2^{n}\right) = n  \ . \]
High minimum entropy guarantees with high probability any adversary will fail to guess the password even after many guesses. However, even minimum entropy is not a great measure of security when the user is managing multiple passwords because it does not consider correlations between passwords. Suppose for example that each user needs two passwords $\left(x_1,x_2\right)$ and again consider two password distributions $D_1$ and $D_2$ redefined below:

\[D_1\left(n\right) = \left(x,x\right)~\text{ with probability $2^{-2n}$ for each $x \in \{0,1\}^{2n}$} \ .\]

\[D_2\left(n\right) = \left(x_1,x_2\right)~\text{ with probability $2^{-2n}$ for each $\left(x_1,x_2\right) \in \{0,1\}^n\times\{0,1\}^n $} \ .\]

The min-entropy of both generators is the same ($2n$). However, $D_1$ provides no security guarantees against a recovery attack --- any adversary who knows $x_2$ can immediately guess $x_1$. However, when the passwords are chosen from $D_2$ an adversary who knows $x_2$ has no advantage in guessing $x_1$.

\section{Economics} \label{appendix:Econ}
In this section we discuss how the parameter $\Guesses{B}$ - our upper bound on the total number of adversary guesses - could be selected. Our upper bound is based on the economic cost of guessing. Guessing is not free! The basic premise is that the adversary will not try more than $\Guesses{\$B}$ guesses to break into an account if his maximum benefit from the attack is $\$B$. The cost of guessing is influenced by several factors including the cost of renting or buying computing equipment (e.g., Cray, GPUs), the cost of electricity to run the computers and the complexity of the cryptographic hash function used to encrypt the password.  The value of $\Guesses{\$B}$ depends greatly on the specific choice of the cryptographic hash function. Table \ref{tab:QUpperBound} shows the values of $\Guesses{\$B}$ we computed for the BCRYPT, SHA1 and MD5 hash functions. 

\begin{table}[t]
\centering
\begin{tabular}{|l|l|l|l|}
\hline 
Benefit (B) & BCRYPT & MD5 & SHA1 \\
\hline 
$\Guesses{B}$ & $B\left(5.155 \times 10^4\right)$ & $B\left(9.1 \times 10^9\right)$ & $B \times 10^{10}$  \\
\hline 
\end{tabular}
\caption{Upper Bound: $\Guesses{B}$ for BCRYPT, MD5 and SHA1}
\label{tab:QUpperBound}
\end{table}

\subsection{Password Storage}
There are many cryptographic hash functions that a company might use (e.g., MD5, SHA1, SHA2, BCRYPT) to store passwords. Some hash functions like BCRYPT \cite{bcrypt} were designed specifically with passwords in mind --- BCRYPT was intentionally designed to be slow to compute (e.g., to limit the power of an  adversary's offline attack). The BCRYPT hash function takes a parameter which allows the programmer to specify how slow the hash computation should be --- we used L12 in our experiments. By contrast, MD5, SHA1 and SHA2 were designed for fast hardware computation. Unfortunately, SHA1 and MD5 are more commonly used to hash passwords \cite{breach:linkedin}. In economic terms, hash functions like BCRYPT increase the adversary's cost of guessing. We use $F_H$ to denote number of times that the hash function $H$ can be computed in one hour on a 1 GHz processor. We estimated $F_H$ experimentally on a Dell Optiplex 960 computer for BCRYPT, MD5 and SHA1 (table \ref{tab:guessingCostsApdx})
 --- as expected the value of $F_H$ is much lower for BCRYPTY than SHA1 and MD5. 

The rainbow table attack can be used to significantly speed up password cracking attempts after the adversary performs some precomputation \cite{rainbowTable}. Rainbow table attacks can be prevented by a practice known as password salting (e.g., instead of storing the cryptographic hash of the password $h(p)$ the a server stores $\paren{h\paren{p,r},r}$ for a random string $r$) \cite{salt}. 

\noindent {\bf Note: }, In reality, many companies do not salt their passwords \cite{breach:linkedin,breach:sony} (in fact some do not even hash them \cite{breach:rockyou}). In this paper, we assume that passwords are stored properly (e.g., salted and hashed), and we use optimistic estimates for $\Guesses{\$B}$ based on the BCRYPT hash function. To justify these decisions we observe that a user could easily ensure that his passwords are salted and encrypted with a slow hash function $f$ (e.g., BCRYPT \cite{bcrypt}) by using $f\paren{U, A_i, p_i}$ as his password for account $i$ - where $U$ is the username and $A_i$ is the name of account $i$. Because the function $f$ is not a secret, its code could be stored locally on any machine being used or publicly on the cloud.

\subsection{Attack Cost and Benefit}
Suppose that company $A_i$ is hacked, and that the usernames and password hashes are stolen by an adversary. We will assume that company A has been following good password storage practices (e.g., company $A_i$ hashes all of their passwords with a strong cryptographic hash function, and company $A_i$ salts all of their password hashes). 
The adversary can purchase any computing equipment he desires (e.g., Cray supercomputer, GPUs, etc) and run any password cracker he wants for as long as he wants. The adversary's primary limitation is money. It costs money to buy all of this equipment, and it costs money to run the equipment. If the adversary dedicates equipment to run a password cracker for several years then the equipment may be obsolete by the time he is finished (depreciation). We define $C_g$ to be the amortized cost per guesses for the adversary. 

\cut{Similarly, we let $p_j$ denote the probability that the adversary succeeds on the j'th guess. Notice that $p_1 \geq p_2 \geq ... \geq p_j$ (otherwise the adversary could reorder his guesses). We let $B_j$ denote the adversaries expected benefit in cracking the password. We discuss factors that influence $B_j$ in section \ref{subsec:BenefitOfGuessing}. 

{\bf Sufficient Condition: } A sufficient condition to prevent password cracking activity is: 
\[ C_g \geq p_i B_A \ . \]
Intuitively, the adversary will only pay for more guesses if the expected benefit $p_i B_A$ exceeds the cost $C_g$ of one more guess. }

\subsection{Cost of Guessing}
Included in the amortized guessing cost are: the price of electricity and the cost of equipment. We estimate $C_g$ by assuming that the adversary rents computing time on Amazon's cloud EC2 \cite{amazonCloudPricing}. This allows us to easily account for factors like energy costs, equipment failure and equipment depreciation. Amazon measures rented computing power in ECUs \cite{amazonCloudPricing} --- ``One EC2 Compute Unit (ECU) provides the equivalent CPU capacity of a 1.0-1.2 GHz 2007 Opteron or 2007 Xeon processor." We use $C_GHz$ to denote the cost of renting a 1 GHz processor for 1 hour on Amazon. We have
\[C_g =  \frac{C_{GHz}}{F_H} \ .\]
Using the Cluster GPU Instance rental option the adversary could rent 33.5 ECU compute units for \$2.10 per hour ( $C_{GHz} = \$.06$).

 Our results are presented in table \ref{tab:guessingCostsApdx}.

\begin{table}[t]
\centering
\begin{tabular}{| c | c | c | }
Hash Function & $F_H$ & $C_{\Guesses{}}$ \\
SHA1 & $\sim 576\times 10^6$ guesses per hour  & $\$1\times 10^{-10}$ \\
MD5 & $\sim 561\times 10^6$ guesses per hour  & $\$1.1\times 10^{-10}$ \\
BCRYPT (L12) & $\sim 31\times 10^3$ guesses per hour  & $\$1.94\times 10^{-5}$ \\
\end{tabular}
\caption{Guessing Costs}
\label{tab:guessingCostsApdx}
\end{table}

\subsection{Benefit} \label{subsec:BenefitOfGuessing}
The benefit $B_j$ of cracking an account $A_j$ is dependent on both the type of account (e.g., banking, e-mail, commerce, social network, leisure) and the adversary's background knowledge about the user (e.g., Does the user reuse passwords? Is the user rich? Is the user a celebrity?). 

Password reuse has a tremendous impact on $B$. An adversary who cracked a user's ESPN account would likely get little benefit --- unless the user reused the password elsewhere. For most non-celebrities, $B_j$ can be upper bounded by the total amount of money that the user has in all of his financial accounts. In fact, this may be a significant overestimate --- even if the user reuses passwords --- because banks are usually successful in reversing large fraudulent transfers \cite{passwordStealing}. Indeed, most cracked passwords sell for between \$4 and \$17 on the black market \cite{passwordBlackMarket}. An adversary might also benefit by exploiting the user's social connections (e.g., tricking the user's friends to wire money). Some user's passwords may also be valuable because have access to valuable information (e.g., celebrity gossip, trade secrets). 

Most users should be safely assume that no adversary will spend more than \$1,000,000 to crack their account even if they reuse passwords. Table \ref{tab:guessingCosts} shows the value of $\Guesses{\$1,000,000}$ for various hash functions. 

\begin{table}[t]
\centering
\begin{tabular}{| c | c | }
Hash Function & $\Guesses{\$1,000,000}$ \\
SHA1 & $10^{16}$\\
MD5 &  $9.1\times 10^{15}$ \\
BCRYPT (L12) & $5.2 \times 10^{10}$    \\
\end{tabular}
\caption{$\Guesses{\$1,000,000}$}
\label{tab:guessingCosts}
\end{table}

\section{Associative Memory and Sufficient Rehearsal Assumptions} \label{appendix:Memory}
The expanding rehearsal assumption makes empirical predictions about long term memory retention (e.g., a user who follows a rehearsal schedule for a cue-association pair will retain that memory for many years). Empirical studies of human memory are often limited in duration due to practical constraints.

The most relevant long term memory study was conducted by Wozniak and Gorzelanczyk \cite{memory:ExpandingRehearsal}. They supervised a group of 7 people who learned 35,000 Polish-English word pairs over 18 months. Their goal was to optimize the intervals between rehearsal of each word pair. They ended up with the following recursive formula 
\[ I(EF,R) = I(EF,R-1)\times OF(EF, R) \ , \]
where  $ I(EF,R)$ denotes the time interval before the $R$'th rehearsal, $EF$ denotes the easiness factor of the particular word pair, and $OF(EF, R)$ is a coefficient matrix which specifies how quickly the intervals grow
\footnote{SuperMemo, a popular commercial memory program \url{http://www.supermemo.com/}, also uses a similar rehearsal schedule.}. The intervals are very similar to those generated by the expanding rehearsal assumption. Our association strength parameter $\AssociationStrength{}$ is similar to the easiness factor $EF$. However, in the expanding rehearsal assumption $OF(EF, R)$ would be a constant that does not vary with $R$. 

Squire tested very long term memory retention by conducting a series of studies over 30 years  \cite{memory:forgetting:squire1989}. To conduct his studies Squire selected a TV show that was canceled after one season, and quizzed participants about the show. It was not surprising that participants in the early studies --- conducted right after the show was canceled --- had the best performance on the quizzes. However, after a couple of years performance dropped to a stable asymptote \cite{memory:forgetting:squire1989}. The fact that participants were able to remember some details about the show after 30 years suggests that it is possible to maintain a cue-association pair in memory without satisfying all of the rehearsal requirements given by our pessimistic constant rehearsal assumption.

\subsection{Squared Rehearsal Assumption}
Anderson and Schooler demonstrated that the availability of a memory is corellated with recency and the pattern of previous exposures (rehearsals) to the item \cite{memory:alternateanderson1991reflections}. Eventually, the following equation was proposed

\[A_i\paren{t} = \sum_{j=1}^n \frac{1}{\sqrt{t-t_j}} \, \]
where $A_i\paren{t}$ denotes the availability of item $i$ in memory at time $t$ and $t_1,\ldots t_n < t$ denote the previous exposures to item $i$ \cite{memory:alternatevan}. In this model the rehearsal schedule $R\paren{\vec{c},j} = j^2$ is sufficient to maintain high availability. To see this consider an arbitrary time $t$ and let $k$ be the integer such that $\paren{k^2 < t \leq (k+1)^2}$. Because $t_k = k^2 < t$ at least $k$ previous rehearsals have occured by time $t$ so

\[A_i\paren{t} = \sum_{j=1}^k \frac{1}{\sqrt{t-t_j}} 
 = \sum_{j=1}^k \frac{1}{\sqrt{t-j^2}} 
= \sum_{j=1}^k \frac{1}{\sqrt{\paren{k+1}^2}} 
\geq \frac{k}{k+1}  \ .\]
\cut{
\begin{eqnarray*}
A_i\paren{t} &=& \sum_{j=1}^k \frac{1}{\sqrt{t-t_j}} \\
 &=& \sum_{j=1}^k \frac{1}{\sqrt{t-j^2}} \\
&=& \sum_{j=1}^k \frac{1}{\sqrt{\paren{k+1}^2}} \\
&\geq& \frac{k}{k+1} \ .
\end{eqnarray*}
}
{\bf Squared Rehearsal Assumption (SQ): } The rehearsal schedule given by $R\paren{\hat{c},i} = i^2 \AssociationStrength{}$ is sufficient to maintain the association $(\hat{c},\hat{a})$.  \\

While SQ is certainly not equivalent to ER it is worth noting that our general conclusions are the same under both memory assumptions. The rehearsal intervals grow with time under both memory assumptions yielding similar usability predictions --- compare Tables \ref{tab:UsabilityOld},\ref{tab:Usability} and \ref{tab:UsabilityAlternate}. The usability predictions are still that (1) \StrongRandomPassword~---though highly secure --- requires any user with infrequently visited accounts to spend a lot of extra time rehearsing passwords, (2) \Lifehacker~requires little effort --- but it is highly insecure, (3) SC-0, which is almost as good as \Lifehacker~from a usability standpoint, provides the user with some provable security guarantees, and (4) SC-1 and SC-2 are reasonably easy to use (except for the Infrequent user) and provide strong provable security guarantees --- though not as strong as \StrongRandomPassword.

While the expanding rehearsal assumption yields fewer rehearsal requirements over the first year, the usability results for \Lifehacker~and  \sharedCues~are even stronger because the intervals {\em initially} grow faster. The usability results are worse for \StrongRandomPassword~because many of the cues are naturally rehearsed with frequency $\lambda = 1/365$ --- in this case most rehearsal requirement will require an extra rehearsal\footnote{The usability results for our occasional user are better than the very active user because the occasional user has fewer sites that a visited with frequency $\lambda = 1/365$. }.

\begin{table}
\centering
\begin{tabular}{| l | l | l | l |l  | l | }
\hline
Schedule/Scheme  & B+D & SC-0 & SC-1& SC-2 & SRI  \\
\hline
Very Active & $\approx 0$ & $\approx 0$ & $2.77$  & $5.88$ & $794.7$ \\
\hline
Typical & $\approx 0$ & $\approx 0$ & $7.086 $   & $12.74$ & $882.8$  \\
\hline 
Occasional & $\approx 0$ & $\approx 0$ & $8.86$& $16.03$ & $719.02$ \\
\hline
Infrequent & $.188$  & $2.08$ & $71.42$ & $125.24$ & $1176.4$  \\
\hline
\end{tabular}
\caption{$E\left[\TotalExtraRehearsals{365}\right]$: Extra Rehearsals over the first year under the Squared Rehearsal Assumption --- $\AssociationStrength{}=1$. \newline B+D: \Lifehacker~~~ \newline SRI: \StrongRandomPassword} 
\label{tab:UsabilityAlternate}
\end{table}

\end{document}